\documentclass[sn-mathphys-num]{sn-jnl}

\usepackage{graphicx}%
\usepackage{multirow}%
\usepackage{amsmath,amssymb,amsfonts}%
\usepackage{amsthm}%
\usepackage{mathrsfs}%
\usepackage[title]{appendix}%
\usepackage{xcolor}%
\usepackage{textcomp}%
\usepackage{manyfoot}%
\usepackage{booktabs}%
\usepackage{algorithm}%
\usepackage{algorithmicx}%
\usepackage{algpseudocode}%
\usepackage{listings}%

\theoremstyle{thmstyleone}%
\theoremstyle{thmstyletwo}%

\theoremstyle{thmstylethree}%

\begin{document}

\title[GLA-GAN]{MRI-to-PET Cross-Modality Translation using Globally \& Locally Aware GAN (GLA-GAN) for Multi-Modal Diagnosis of Alzheimer’s Disease}


\author[1]{\fnm{Apoorva} \sur{Sikka}}
\equalcont{These authors contributed equally to this work.}
\author[2]{\fnm{Skand} \sur{Peri}}
\equalcont{These authors contributed equally to this work.}
\author[1]{\fnm{Jitender Singh} \sur{Virk}}
\author[1]{\fnm{Usma} \sur{Niyaz}}
\author*[1]{\fnm{Deepti R.} \sur{Bathula}}\email{bathula@iitrpr.ac.in}

\affil[1]{\orgdiv{Department of Computer Science and Engineering}, \orgname{Indian Institute of Technology Ropar}, \state{Punjab}, \country{India}}

\affil[2]{\orgname{Oregon State University},\city{Corvallis}, \postcode{OR 97331}, \country{United State}}




\abstract{Medical imaging datasets are inherently high dimensional, with large variability and low sample sizes that limit the effectiveness of deep learning algorithms. Recently, generative adversarial networks (GANs) with the ability to synthesize realist images have shown great potential as an alternative to standard data augmentation techniques. Our work focuses on the cross-modality synthesis of fluorodeoxyglucose~(FDG) Positron Emission Tomography~(PET) scans from structural Magnetic Resonance~(MR) images using generative models to facilitate multi-modal diagnosis of Alzheimer’s disease (AD). Specifically, we propose a novel end-to-end, globally and locally aware image-to-image translation GAN (GLA-GAN) with a multi-path architecture that enforces global structural integrity and fidelity to local details. We further supplement the standard adversarial loss with voxel-level intensity, multi-scale structural similarity (MS-SSIM), and region-of-interest (ROI) based loss components that reduce reconstruction error, enforce structural consistency at different scales, and perceive a variation in regional sensitivity to AD, respectively. Experimental results demonstrate that our GLA-GAN not only generates synthesized FDG-PET scans with enhanced image quality but also superior clinical utility in improving AD diagnosis compared to state-of-the-art models. Finally, we attempt to interpret some of the internal units of the GAN that are closely related to this specific cross-modality generation task.}




\maketitle

\section{Introduction}\label{sec1}

Alzheimer's disease~(AD) is a chronic neuro-degenerative disorder that interferes with memory, thinking, and behavior. It is a progressive disease where symptoms gradually worsen, making early diagnosis crucial. In addition to clinical and behavioral observations, the diagnosis of AD is based on neuroimaging assessments.  Recently, efforts have been directed towards multi-modal image analysis to identify reliable biomarkers that can aid in the accurate diagnosis of AD \cite{multimodal}. As multiple imaging modalities capture complementary information related to different aspects of the disease, their synergistic fusion increases the efficacy of diagnosis.  While gray matter atrophy and ventricular enlargement observed in magnetic resonance imaging~(MRI) are established markers for pathology, patterns of neuronal uptake and cerebral distribution of fluorodeoxyglucose~(FDG) in positron emission tomography~(PET) is also an essential discriminating factor for AD \cite{fdgAD}. Consequently, the joint analysis of MRI and PET is now a standard evaluation for AD \cite{johnson2012brain, zhang2017pet, CNNmulti}. However, acquiring different modality scans for all patients is not always feasible due to high cost, increased risk of radiation exposure, and limited access to imaging facilities. While MRI scanners are quite pervasive, PET technology is not yet common. Our work focuses on cross-modality estimation of FDG-PET scans from MR images using generative models. We further assess the quality of the synthesized PET images based on their potential to diagnose AD.

With recent advances in machine learning, several studies have focused on the task of image-to-image translation which involves transforming images from one domain to another. \cite{huynh2015estimating} used a structured random forest-based model along with an auto-context model to translate MRI patches into computed tomography~(CT) for attenuation correction in PET. Similarly, a regression forest-based framework was developed in \cite{lowdose} to predict a standard-dose brain FDG-PET image from a low-dose PET image and its corresponding MRI. \cite{jog2014random} has used regression forests to reconstruct a FLAIR image given the corresponding T1-weighted, T2-weighted, and PD-weighted images of the same subject.

In the past few years, Generative Adversarial Networks~(GAN)~ \cite{GAN} have achieved remarkable success due to their versatility in synthetic image generation with potential applications across various domains \cite{DCGAN_ICLR, stackgan}. GANs are being utilized extensively in medical image analysis to perform a wide range of tasks, including denoising, reconstruction, segmentation, classification etc. \cite{denoising, mrtoct, frid2018gan, han2018gan, han2018spine}. The ability of GANs to generate diverse and realistic image samples makes them especially appealing to the medical domain that suffers from a lack of large, annotated datasets. Before GANs, other generative models like Autoencoders \cite{sevetlidis2016whole} and Variational Autoencoders (VAE) \cite{kingma2013auto} were also utilized. Several variants of GANs have been proposed to tackle cross-modality estimation or data augmentation in the medical domain. Conditional GANs, where image generation is conditioned on external information, have been used in many cross-modality image translation tasks such as MRI to CT \cite{ACM-GAN}, CT to MRI \cite{Jin2019DeepCT}, etc.  To reduce radiation dose, \cite{ACM-GAN} employed context-aware GANs to predict CT scans from MR scans using an image gradient difference loss function. Although MR and CT represent different imaging modalities, they both capture details of anatomical structures.

A more challenging cross-modality synthesis task involves predicting functional scans from their corresponding structural scans. In \cite{cttopet}, PET images are estimated from CT scans using a mixed Fully Convolutional Network~(FCN) and a conditional GAN approach where the output of FCN is fed in addition to the CT scan as an input using only pixel-wise and adversarial loss. A sketcher-refiner-based adversarial training has also been employed to predict PET-derived demyelination from multi-modal MRI, including measures derived from diffusion tensor imaging (DTI) \cite{wei2019predicting}. \cite{patch} used a patch-based convolutional neural network ~(CNN) to capture non-linear mapping between PET and MRI. \cite{multimodal} attempted to find a relationship between brain atrophy (MRI) and hypometabolism (PET) using a computationally tractable formulation of partial least squares regression (PLSR). Their main aim was to explore non-local intensity correlations between these modalities through PLSR. \cite{MRI2PET_MICCAI} proposed a global 3D U-Net-based formulation to predict the PET from MR. Recently, \cite{mritopet} proposed an image-to-image translation framework based on 3D cycle-consistent GAN~(3D-cGAN) for synthesizing PET images from MRI. Cycle GANs were initially introduced to train unsupervised image translation models using the GAN architecture. DUAL-GLOW, by \cite{Glow}, is another MRI-to-PET modality transfer technique based on the flow-based generative model \cite{kingma2018glow}. This formulation is based on two flow-based invertible networks and a relation network that maps the latent spaces to each other. 

As observed from previous studies \cite{patch}, a purely local approach is not capable of capturing the true relationship between two complex modalities. This knowledge necessitates using global approaches can capture non-local, non-linear correlations. However, previous attempts \cite{MRI2PET_MICCAI} also indicate that an entirely global approach fails to represent finer structures and details effectively. We hypothesize that the quality of the synthesized PET scans can be improved by exploiting both global and local contexts simultaneously within a GAN architecture. Furthermore, the majority of the previous studies used per-pixel loss functions that tend to generate blurry target images with loss of details as well as cause artifacts~\cite{perceptual}. Consequently, we propose to use alternative loss functions that tackle these issues.  As SSIM is widely used to evaluate the quality of reconstructed images, we use a multi-scale version of SSIM (MS-SSIM) as an additional objective function to assess image quality and ensure structural consistency at different scales \cite{ms-ssim}. In addition, as Alzheimer-related literature suggests significant variation in regional sensitivity to the disease progression, we propose to enforce local contextual integrity using region-wise loss based on anatomical delineation. The main contributions of our work are the following:

\begin{itemize}
\item We propose a novel end-to-end, globally and locally aware image-to-image translation GAN (GLA-GAN) with a multi-path architecture that ensures both global structural integrity and fidelity to local details.
\item We supplement the standard adversarial and voxel-level loss terms with objective functions that ensure structural consistency over multiple resolutions as well as local contextual integrity based on anatomical brain regions.
\item We provide an extensive evaluation of our proposed GLA-GAN against state-of-the-art techniques using numerous quality assessment metrics. Additionally, we demonstrate the potential of synthesized PET scans to improve multi-modal Alzheimer’s disease classification accuracy.
\item Finally, we attempt to gain a better understanding by identifying, visualizing, and interpreting internal GAN features. We present some units that are closely related to both common and differential FDG uptake patterns in PET images. 
\end{itemize}
\section{Methods} 

GANs have been used extensively for generating images for a variety of tasks in computer vision. GANs have two competing neural networks, a generator~($G$) and a discriminator~($D$), that are trained simultaneously in a min-max fashion. The goal of the generator is to create realistic-looking synthetic images to fool the discriminator while the discriminator tries to distinguish between the real and synthetic samples. This competition helps GANs to analyze, capture and mimic the variations within the dataset effectively. Here we use a variant of GAN where the source modality image as an input controls and guides the generator to estimate the target modality image.


Let $\{X_{i=1}^{N}\}$ represent MRI scans and $\{Y_{i=1}^{N}\}$ represent PET scans of $N$ subjects in the training set with dimensions $x_1 \times x_2 \times x_3$ and $y_1 \times y_2 \times y_3$ respectively. Hence, $\{X_{i},Y_{i}\}$ represents the $i^{th}$ subject-specific source-target paired data. The goal is to learn a data-driven mapping function $f: X \longrightarrow Y$ that accurately predicts the target PET image ($Y$) for any new input MR image ($X$). Unlike natural image translation, the mapping between multi-modal medical images can be very complex and highly non-linear. More specifically, as MRI and PET represent structural and functional images of the brain, respectively, they lack local spatial consistency between modalities. This knowledge necessitates the use of global approaches that can capture non-local correlations. However, a purely global approach might not be capable of representing finer structures and details effectively. Our proposed solution combines locally extracted features with global attributes to enhance PET synthesis.


 
\subsection{Globally \& Locally Aware GAN (GLA-GAN)}
\label{sec:Netarch}

\begin{figure}[htbp]
  \centering
  \includegraphics[width=\textwidth]{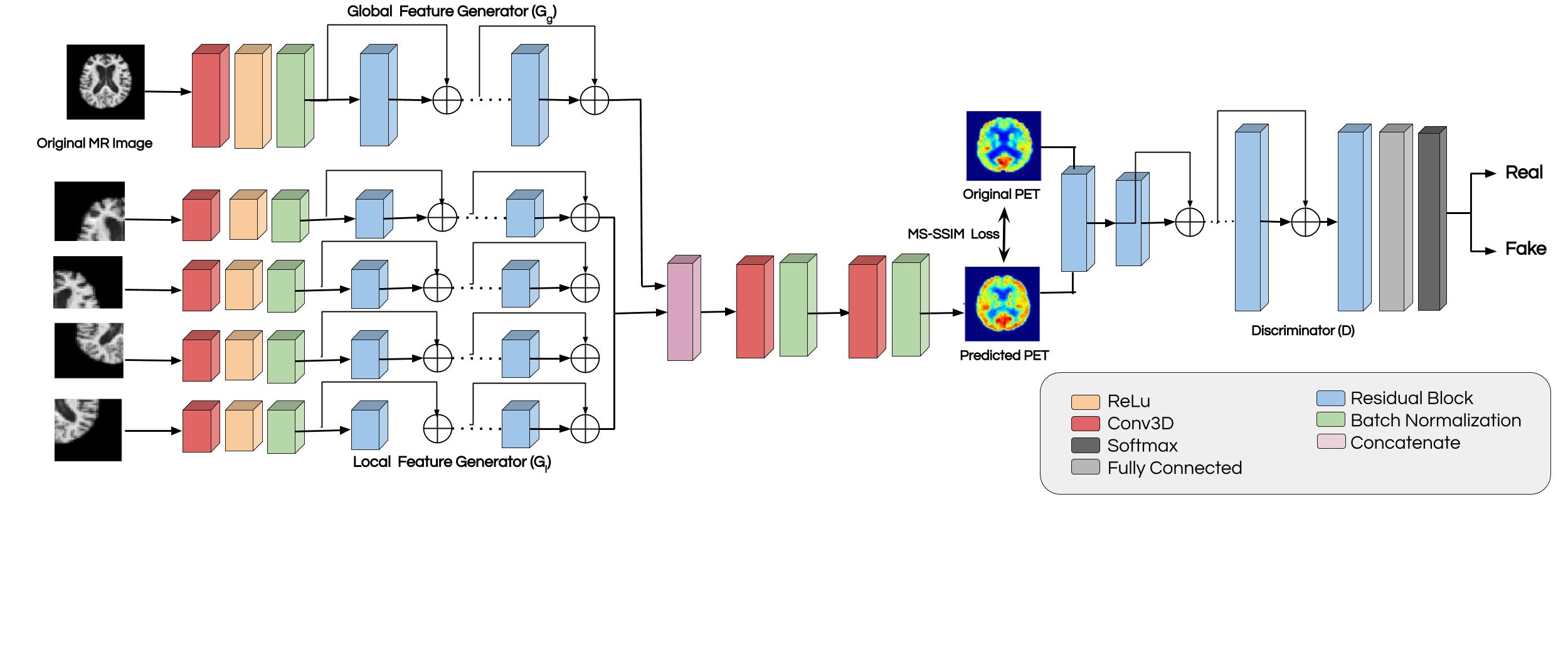}
  \caption{Proposed globally and locally aware image-to-image translation GAN (GLA-GAN) architecture for cross-modal PET estimation from MRI using $L_1$, MS-SSIM, and ROI-based objective function. [Best viewed in color]}
  \label{fig:intro}
\end{figure}

To this extent, we developed a GAN-based framework with multi-path architecture that discerns both global structures and local textures. Figure \ref{fig:intro} provides an overview of the proposed multi-path GAN framework for MRI-to-PET cross-modality transfer. Here, we provide details of the architecture, including a description of the generator and discriminator, along with the loss formulation of our method.

\subsubsection{Generator}
The generator $G$ of our GLA-GAN is a combination of multiple sub-networks. It consists of two modules namely, a \textit{global module} ($G_g$) and a \textit{local module} ($G_l$). The global module is a generator network that maps the complete MR image of dimensions $64 \times 64 \times 64$ to a latent feature space $f_g$ via a set of residual blocks \cite{resnet}. For an input image $X$, the output of global network is $G_g(X) \in R^{64 \times64 \times 64 \times f_g}$. The local module is a set of $K$ generator networks which work independently on $K$ disjoint patches that cover the whole MR image. Each local network maps an input patch ($X_k$) to a feature vector $f_l^k$ and outputs $G_l(X_k) \in R^{32 \times32 \times 64 \times f_l^k}$. The $k$ obtained patches are fused together to form $64 \times 64 \times 64$ before passing to the next layer. The underlying architecture of $G_g$ and each of $G_l^k$ is a set of four residual blocks with batch normalization \cite{bn} and rectified linear unit (ReLU) activation \cite{Nair2010RectifiedLU}. The two sets of features ($f_g$ and $f_l^k$ where $k=1$ to $K$) are directly concatenated channel-wise which maintains their spatial order and passed through two convolutional layers followed by sigmoid activation to regress the intensity values of the target PET image.


\textbf{ResNet Block} - In general, the architecture of most generator modules is a 10-layer ResNet proposed by \cite{resnet}. The basic idea behind ResNet is the introduction of skip connections between layers of a convolution neural network to preserve the input from previous layers. This approach mitigates the gradient degradation problem that is caused by an increase in the number of layers in a network and avoids any loss of information. Our proposed generator network has four residual blocks alongside four convolution layers. The number of convolutional filters is $64$, $64$, $128$, and $64$ respectively with kernel size of $3 \times 3$ and stride 1. On the lines of \cite{DCGAN_ICLR}, we also use batch normalization \cite{bn} after every convolutional layer along with LeakyReLu as the activation function.



\subsubsection{Discriminator}

The discriminator module, $D$, is a simple classification network that forces the generator to produce realistic and coherent PET images. This is achieved by coercing the generator to learn the underlying data distribution by inferring if the generated PET images are real or fake. The discriminator network is a CNN with six convolution layers followed by batch normalization and leaky ReLU activation function for the first five layers and a sigmoid activation after the last layer. The kernel size is $4 \times 4 \times 4$ for the first four layers and $2 \times 2 \times 2$ for the next two layers, and the numbers of the filters in each layer are 64, 128, 256, 512, 64 and 1 respectively.

\subsection{Synthesis Loss function}

The objective function used in our GLA-GAN framework is a combination of multiple loss terms that coerce the network towards the most feasible distribution representing a real dataset.

\subsubsection{Adversarial Loss}
The loss function for the generator is implicit as it learns based on discriminators' performance during training. On the other hand, the discriminator is updated by minimizing the adversarial loss function as given below:

\begin{equation}
    \mathcal{L}_{G}(D,G) = \mathbb{E}_{y \sim Y}(y)[\log D(y)] + \mathbb{E}_{{x} \sim X}[\log (1 - D(G(x)))]
\end{equation}

The idea is to minimize this objective function so that $D$ be as confused as possible to distinguish between real and generated PET images to help $G$ generate increasingly realistic PET scans. However, due to the intricate nature of medical images, only adversarial loss cannot be sufficient to generate high-quality scans that have the same uptake as real PET scans. To mitigate this, we propose to incorporate additional terms in the loss function that penalizes loss of detail to enhance the quality of synthesized scans as described below.

\subsubsection{Voxel-wise $L_1$ Reconstruction Loss}
Least absolute deviations ($L_1$) or voxel-wise reconstruction loss is a standard loss function defined as:

\begin{equation}
    \mathcal{L}_{L_1} = \frac{1}{n}{\sum_{i=1}^{n}{|y_i - \hat{y_i}|}}
\end{equation}


\noindent where $y_{i}$ and $\hat{y_{i}}$ are the intensity values of the voxels in the estimated and real PET scans. By minimizing the sum of all the absolute differences across voxels, $L_1$ loss trains the network to understand the differences between images at a voxel level.
 
\subsubsection{Multi-Scale SSIM Loss}
\label{sec:SSIM}
In addition to adversarial loss, we wanted to enforce the structural similarity in images that encourages the structure of the generated PET to be similar to that of the ground truth PET. Structural similarity index~(SSIM) is one such metric that is used to evaluate the quality of images in image processing algorithms. Single-scale SSIM is computed as follows:

\begin{equation}
    \mathrm{SSIM}(x,y) = [l(x,y)]^\alpha . [c(x,y)]^{\beta} [s(x,y)]^{\gamma}
\end{equation}

\noindent where $l(x,y)$, $c(x,y)$ and $s(x,y)$ are luminance, contrast, and structure measures of the image, and $\alpha$, $\beta$, and $\gamma$ are weight parameters that assign importance of each of the measures. When $\alpha=\beta=\gamma=1$, the above formulation can be reduced to:

\begin{equation}
\mathrm{SSIM}(x,y) = \frac{(2\mu_x\mu_y + C_1) (2 \sigma _{xy} + C_2)} 
    {(\mu_x^2 + \mu_y^2+C_1) (\sigma_x^2 + \sigma_y^2+C_2)}
\end{equation}

\noindent where $x$ is the estimated PET and $y$ is the ground truth PET. $\mu_i$ is the mean of image $i$, $\sigma_i$ is the variance of image $i$ and $\sigma _{xy}$ is the co-variance of images $x$ and $y$. $C_{1}$ and $C_{2}$ are empirically found constants in order to best perceive the structure of the estimated image with respect to the ground truth image.

Given that SSIM is differentiable, it can be easily used as a cost function to induce sensitivity to changes in the structure. However, SSIM is only sensitive to the scale at which local structure is analyzed. Hence, \cite{ms-ssim} proposed a multi-scale variant known as multi-scale structural similarity index~(MS-SSIM) that computes SSIM at multiple scales. In order to compute MS-SSIM for $S$ scales, the image is downsampled by a factor of $2$ for $S-1$ times and is combined using the following formulation:

\begin{equation}
    \mathrm{MS\mbox{-}SSIM}(x,y) = w(s) \times [l(x,y)] . \prod_{j=1}^{S} [c_j(x,y)]^{\beta_j} [s_j(x,y)]^{\gamma_j}
\end{equation}

\begin{equation}
    L_{\mathrm{MS-SSIM}} = 1 - \mathrm{MS\mbox{-}SSIM}
\end{equation}

\noindent where $l(x,y)$, $c_j(x,y)$ and $s_j(x,y)$ are the luminance, contrast, and structural components of the image and $w(s)$ are the weight factor for the scale $S$. In all our experiments, we use empirically determined parameter values: three levels of resolution ($S=3$) and weights as $w(1)=0.0448$, $w(2)=0.2856$, and $w(3)=0.3001$.  This approach ensures structural consistency across multiple resolutions and thus improves the visual quotient of the generated PET scans.

\subsubsection{Region of Interest~(ROI) Loss}

While the $L_1$ loss ensures voxel level fidelity of intensity values, the perceptual similarity metric facilities global structural similarity at multiple scales. However, both of these terms represent summary statistics that consider the image as a whole. As past studies suggest that there is significant variation in regional sensitivity to Alzheimer’s across subjects \cite{Gorden2014}, we propose to enforce local contextual integrity using region-wise loss based on anatomical delineation. We hypothesize that supplementing $L_1$ and MS-SSIM loss with region-based loss can help capture discriminating features between the two classes and thus help GAN to reliably learn a bimodal distribution. Consequently, we used the Automated Anatomical Labeling~(AAL) atlas to segment each MRI scan into 116 ROIs and calculated the mean intensity of all 116 ROIs for both real and generated PET data. These average intensities were used as feature vectors to compute mean-squared ROI loss as follows:

\begin{equation}
 \mathcal{L}_{ROI} = \frac{1}{R} \sum_{i=1}^{R} [\frac{1}{N_R} \sum_{j=1}^{N_R} (y_{ij} - \hat{y}_{ij})]^2
\end{equation}

\noindent where $R$ is number of ROIs, $N_R$ is number of voxels in ROI $R$,
$y_{ij}$ and $\hat{y}_{ij}$ are the true and predicted voxel intensity values from ground truth and synthesized PET scans, respectively.

\subsubsection{Combined Objective Function}

All the modules or sub-networks in the framework are jointly optimized using the following objective function, specially curated for MRI-to-PET cross-modality estimation:

\begin{equation}
    \mathcal{L} = \lambda_{G}\mathcal{L}_{G} + \lambda_{MS-SSIM}\mathcal{L}_{MS-SSIM} +\lambda_{L1}\mathcal{L}_{L1} + \lambda_{ROI}\mathcal{L}_{ROI}
\end{equation}

\noindent where $\lambda_G$, $\lambda_{MS-SSIM}$, $\lambda_{L1}$ and $\lambda_{ROI}$ are weighing factors for GAN, perceptual, voxel-wise, and ROI-based loss terms respectively

\section{Experiments}
\subsection{Dataset}
We conducted all our experiments on the Alzheimer's Disease Neuroimaging Initiative (ADNI) database ~(adni.loni.usc.edu). The ADNI is a longitudinal study designed to develop clinical, imaging, genetic, and biochemical biomarkers for the early detection and tracking of Alzheimer's disease (AD) \cite{adni}. Subjects in the dataset are categorized as either cognitively Normal (CN) or Alzheimer's disease (AD). We collected a total of 402 samples from ADNI-1 \& ADNI-2 having both prepossessed MRI and FDG-PET modalities~(CN-241, AD-161). Further, we collected 179 samples from ADNI-1 \& ADNI-2, which had only MRI scans~(CN-83, AD-79). We refer to the first subset with both modalities as \textbf{Paired} and the fusion of these two subsets as \textbf{Incomplete} datasets, respectively. Lastly, multiple sets of \textbf{Complete} dataset are created by adding synthesized versions of missing PET scans to the Incomplete dataset using various models. 

Details of the image acquisition protocols along with initial preprocessing for ADNI-1 and ADNI-2 scans are described in \cite{jack2010hypothetical, jack2008alzheimer} respectively. A standard sequence of pre-processing steps has been employed to prepare the data for analysis. Firstly, MR images were Anterior Commissure (AC)-Posterior Commissure (PC) corrected using SPM12 \cite{parce}. Then skull stripping of MR images was performed using ROBEX \cite{robex} followed by intensity normalization and linear alignment to the MNI template using FSL package \cite{fsl}. Subsequently, PET images were aligned to MRI using the same linear registration. This brings both MRI and PET images to Montreal Neurological Institute (MNI) space with size $109 \times 90 \times 90$ which were further resized to $64 \times 64 \times 64$. Sub-sampling was performed to accommodate network size to fit into the memory. Finally, MRI images were segmented into gray matter (GM), white matter (WM), and cerebrospinal fluid (CSF) using FSL. The GM and WM are further segmented into ROIs according to the AAL atlas, and each ROI's mean intensity was used as a feature for AD classification.

\subsection{Experimental Setup}

All the experiments were conducted using Pytorch \cite{paszke2017automatic}. We trained the network end-to-end  with a batch size of 4, using Adam optimizer \cite{adam} with a momentum of 0.1 and a learning rate of 0.00002.


\subsubsection{Evaluation Metrics}
\label{sec:evaluation}

Following is a brief description of the metrics used to evaluate the quality of the synthesized images.

\textbf{Mean Absolute Error}  is a commonly used metric for prediction problems to measure the magnitude of voxel-wise intensity differences. It is an arithmetic average of the absolute differences between the estimated and true intensity values. It is computed as follows: 

        \begin{equation}
        MAE = \frac{1}{n} \sum_{i=1}^{n}{|y_i - \hat{y}_i|}
        \end{equation}
    
\noindent where $y_{i}$ and $\hat{y}_{i}$ are the intensity values of the voxels in the predicted and ground truth PET scans, respectively.

\textbf{Peak Signal-to-Noise Ratio~(PSNR)} is most commonly used to measure the quality of reconstruction algorithms. The metric is a ratio of the maximum power of a signal and the power of distorting noise that affects the fidelity of its representation. It is generally expressed in terms of the logarithmic decibel scale. It is computed as follows:
    

    \begin{equation}
    PSNR= 10 \; log_{10} \Bigg( \frac{MAX^2}{MSE} \Bigg)
    \end{equation}

\noindent, where $MAX$ is the maximum possible intensity of the image and $MSE$, is the mean squared error between the estimated and ground truth PET image. A higher value of PSNR is indicative of better synthesis quality.

\textbf{Structure Similarity Index~(SSIM)} compares the similarity in structures of the two images instead of comparing individual voxel values as explained in the section. \ref{sec:SSIM}

\subsubsection{Comparison with State-Of-The-Art (SOTA) methods}
\label{sec:comparison}

We investigated the efficacy of our proposed methodologies by comparing them to different approaches developed specifically to address MRI-to-PET modality transfer.  Although most related studies used the ADNI dataset, the sample subset chosen to conduct experiments varied across them. To ensure a fair comparison, we accurately re-implemented three of the state-of-the-art networks: 3D U-Net based autoencoder \cite{MRI2PET_MICCAI}, Conditional GAN \cite{conditional} and Cycle-GAN \cite{mritopet}.  A comprehensive evaluation of all the methods is performed using both qualitative and quantitative assessment tools. We used the standard \textit{k}-fold cross-validation with 5 folds to evaluate the models. 


\subsubsection{Alzheimer's Classification}
\label{sec:ADclass}
\begin{figure*}[h!]
    \centering
   
        \includegraphics[width=\textwidth,keepaspectratio]{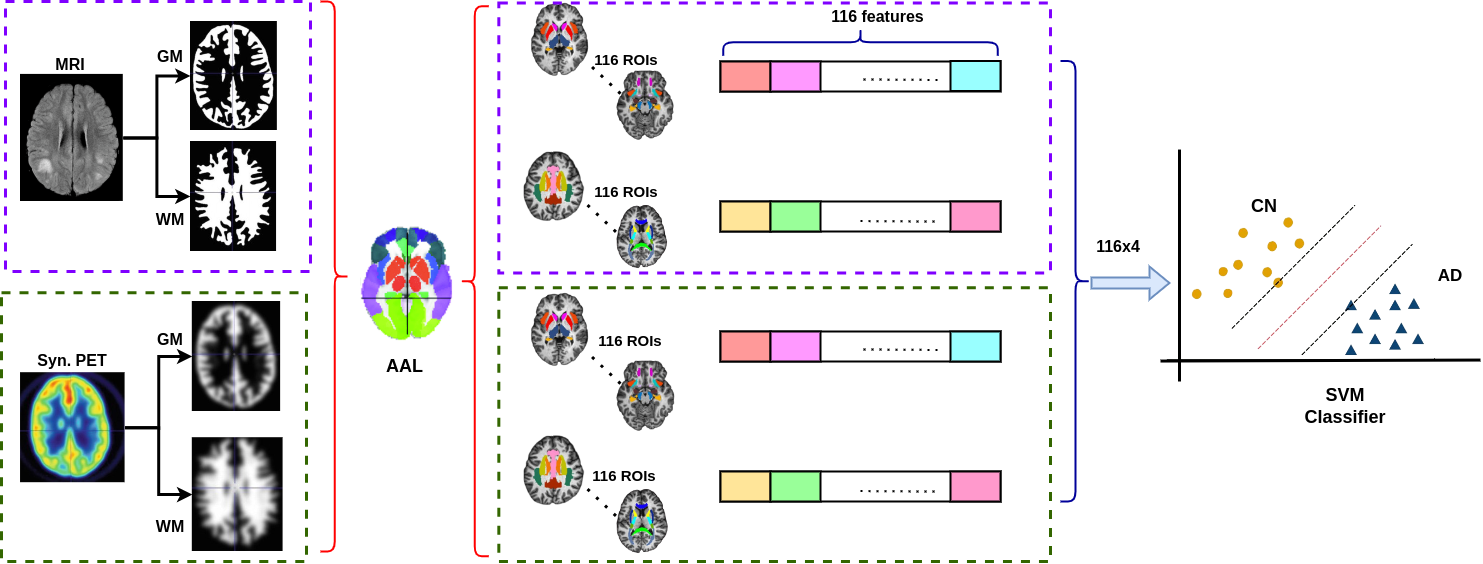}
      
   \caption{Classification Model: Using features extracted from MR images and their corresponding synthetic PET images 
produced through various generative methods [Best viewed in color]}
\label{fig:classify}
\end{figure*}

In addition to the quality of the synthesized PET scans, we showcase the potential of the generated PET scan to facilitate binary Alzheimer's disease classification~(AD vs Normal). Towards this end, we used GM and WM segmentation first and did the Automated Anatomical Labeling~(AAL) atlas to segment each MR image into 116 ROIs and calculated the mean intensity of each ROI. These average intensities taken from both modalities lead to a feature vector of size $116 * 4$. This final feature vector was used to train a support vector machine~(SVM) using a radial basis function~(RBF) kernel for multi-modality AD classification as shown in Figure \ref{fig:classify}. We use several standard classification performance metrics such as accuracy (ACC), sensitivity (SENS), specificity (SPEC), Area under the curve (AUC), Matthews correlation coefficient~(MCC), and F1-score to evaluate the models.

\section{Results \& Discussion}
We demonstrate the merits of our proposed model based on the quality of synthesized PET scans and their potential for accurately diagnosing AD. Additionally, we evaluate the effectiveness of different modules of the multi-path GAN architecture and illustrate the benefits of various loss components through ablation studies. We have compared our method only with other deep learning variants as they proved to be better than traditional ML methods. Finally, we visualize and interpret some internal GAN units in the context of this particular generation task.

\begin{figure*}[htb!]
    \centering
   
        \includegraphics[width=\linewidth]{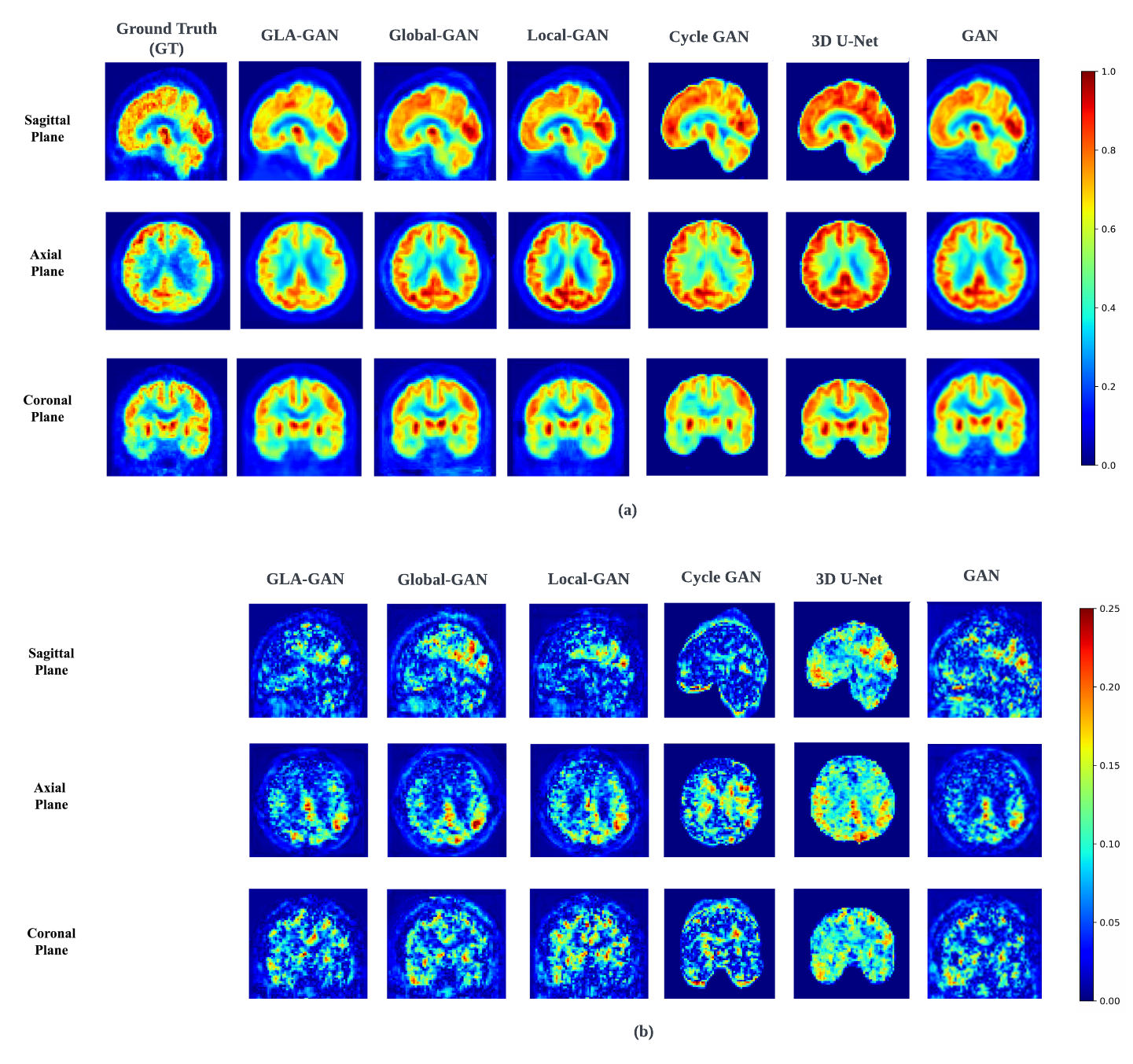}
      
   \caption{Qualitative comparison of PET scans for Alzheimer’s disease (AD) sample: synthesized using GLA-GAN, Global-GAN, Local-GAN, 3D U-Net, Cycle GAN, and GAN. Ground truth, estimated PET scan, and error maps corresponding to each model are presented in axial, coronal, and sagittal viewse. [Best viewed in color]}
\label{fig:result1}
\end{figure*}
\subsection{PET Synthesis}
\label{sec:PETpred}

\begin{figure*}[htb!]
    \centering
   
        \includegraphics[width=\linewidth]{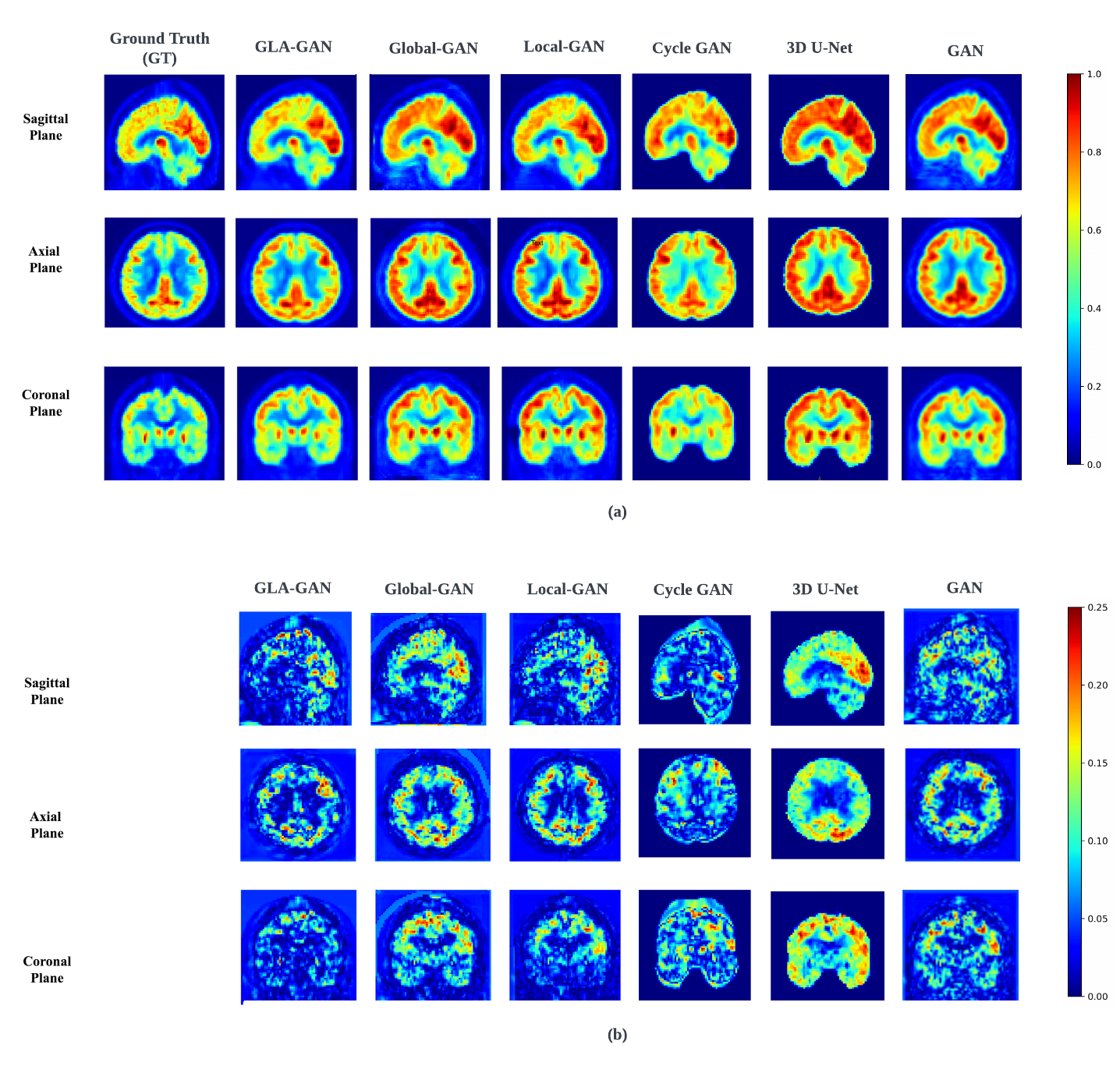}
      
   \caption{Qualitative comparison of PET scans for Control (CN) sample: Synthesized using GLA-GAN, Global-GAN, Local-GAN, 3D U-Net, Cycle GAN, and GAN. Ground truth, estimated PET scan, and error maps corresponding to each model are presented in axial, coronal, and sagittal views. [Best viewed in color]}
\label{fig:result2}
\end{figure*}

Firstly, we demonstrate the significance of simultaneously incorporating global and local contextual information through parallel modules in the GLA-GAN framework. To this end, we compare three variants of the multi-path architecture: (i) Global-GAN, (ii) Local-GAN, and (iii) Global-Local GAN (GLA-GAN). Figure \ref{fig:result1} and Figure \ref{fig:result2} (a) and (b) show qualitative visualizations of PET scans synthesized using these three models along three planes~(axial, sagittal, and coronal) for an AD and CN sample, respectively. In addition to the generated images, error maps with respect to the ground truth are also presented. Although not readily apparent, there are subtle differences in the estimates generated by the individual Global and Local only modules. A closer observation reveals that while the local module is better at capturing intricate details like the shape of ventricles, the global module achieves a better estimate of the FDG uptake intensity. However, it is clearly evident from both synthesized images and error maps that the hybrid model most accurately predicts the FDG uptake patterns for both samples (AD and CN).


Next, we benchmarked our proposed GLA-GAN model against three state-of-the-art models: (i) 3D U-NET based autoencoder \cite{MRI2PET_MICCAI}, (ii) GAN  (L1 + MS-SSIM)   and (iii) Cycle GAN \cite{mritopet}. As Cycle GAN is based on a global transformation framework, it is able to discern general variations in uptake patterns of AD and CN classes. However, it fails to capture local contextual details, as evident from the discrepancy in the shape and uptake in ventricles depicted in both the axial and coronal planes compared to the ground truth PET scan. On the other hand, the 3D U-NET produced very generic and blurry estimates of the real PET scans. Compared to these global methods, the results clearly indicate the superiority of our GLA-GAN model in accurately synthesizing PET scans without losing details.  Additionally, a quantitative evaluation of these models based on three quality assessment metrics is provided in Table \ref{table:qual}. The GLA-GAN framework provides either on-par (statistically insignificant difference) or better performance than competing methods across these metrics. While it provides the lowest MAE, there was no significant difference for PSNR and SSIM ($p = 0.1690$ and $p = 0.1363$ using paired, two-tailed T-test) compared to the best-performing model. \\

\begin{table}[!htbp]
\begin{tabular}{|c|c|c|c|}
\hline
\textbf{Method}                                                          & \textbf{SSIM} ($\uparrow$)            & \textbf{PSNR}  ($\uparrow$)          & \textbf{MAE} ($\downarrow$)\\ \hline
3D U-Net \cite{MRI2PET_MICCAI}                                                           &96.496 $\pm$ 0.010          & 28.928 $\pm$ 0.260         & 0.015 $\pm$ 0.238\\ \hline

C-GAN   \cite{conditional}                           & \textbf{96.974} $\pm$
\textbf{0.181} & \textbf{29.500} $\pm$ \textbf{0.019}          & 0.0144 $\pm$ 0.001        \\ \hline

Cycle GAN   \cite{mritopet}                                                    & 96.148 $\pm$ 0.004          &23.194 $\pm$ 0.220           & 0.0142 $\pm$ 0.005  \\ \hline

Local-GAN & 96.150 $\pm$ 0.049 & 28.525 $\pm$ 0.077 & 0.0157 $\pm$ 0.001 \\ \hline

 Global-GAN & 95.810 $\pm$ 0.012 & 28.268 $\pm$ 0.036 & 0.0162 $\pm$ 0.001 \\ \hline

GLA-GAN & 96.882 $\pm$ 0.260$^\dag$ & 29.326 $\pm$ 0.060$^\dag$ & \textbf{0.0137} $\pm$ \textbf{0.001}
 \\ \hline
\end{tabular}
\caption{Comparison of PET synthesis quality using different metrics. SSIM - Structural Similarity Index, PSNR - Peak Signal-to-Noise Ratio, MAE - Mean Absolute Error. Results are shown as mean $\pm$ standard deviation. (Bold and $^\dag$ represent the first and second best, respectively) }
\label{table:qual}
\end{table}

\subsection{Comparison with SOTA Generative Models}

Table \ref{table:sota} presents the comparative performance of our proposed GLA-GAN model against state-of-the-art methods, evaluated using SSIM, PSNR, and MAE. GLA-GAN achieves the highest SSIM (96.88 ± 0.260), signifying its superior ability to maintain structural similarity while delivering the lowest MAE (0.014 ± 0.001), indicating highly accurate predictions with minimal error. Although its PSNR (29.32 ± 0.260) is higher than some methods, it aligns well with acceptable reconstruction quality standards. Compared to models like U-Net and GAN-based approaches, which perform well in SSIM but show higher MAE, and BPGAN, which achieves lower MAE but struggles with SSIM, our GLA-GAN demonstrates a balanced performance. These results highlight the robustness of our model in handling complex data, delivering both high-quality reconstructions and precise outputs across diverse scenarios.
\begin{table}[!htbp]
\begin{tabular}{|c|c|c|c|}
\hline
\textbf{Method} & \textbf{SSIM} ($\uparrow$)             & \textbf{PSNR}  ($\uparrow$)          & \textbf{MAE} ($\downarrow$) \\ \hline
  
 Wasserstein GAN with Pix2Pix \cite{WANG2020e775}   & 72.40 $\pm$ 0.035   & {69.20} $\pm$ {1.350}   & 0.069 $\pm$ 0.005 \\ \hline

  Pix2Pix \cite{pip2pix}     & 93.00 $\pm$ 0.010   & \textbf{73.94} $\pm$ \textbf{1.520}   & - \\ \hline

    DUAL-GLOW \cite{Sun2019DUALGLOWCF}  & 89.80 $\pm$ 0.060    & 29.56 $\pm$ 2.660  & - \\ \hline

  BMGAN \cite{gen_gan}   & 89.00 $\pm$ 0.080 & {27.88} $\pm$ {1.170} & 25.34 $\pm$ 7.840 \\ \hline

   Adversarial U-Net \cite{gen_adve}  & -  & 24.610 $\pm$ 1.615 & 34.925 $\pm$ 8.784 \\ \hline
   BPGAN \cite{bpgan}  &  70.88 & 26.46 & 0.034 \\ \hline
GLA-GAN    & \textbf{96.88} $\pm$ \textbf{0.260}   & 29.32 $\pm$ 0.260   & \textbf{0.014} $\pm$ \textbf{0.001}\\ \hline
\end{tabular}
\caption{Comparison of PET synthesis quality with state-of-the-art generative models using different metrics. SSIM - Structural Similarity Index, PSNR - Peak Signal-to-Noise Ratio, MAE - Mean Absolute Error. Results are shown as mean $\pm$ standard deviation. (Bold and $^\dag$ represent the first and second best, respectively)}

\label{table:sota}
\end{table}
\subsection{AD Classification}

In addition to the quality of images, the potential clinical utility of the generated scans is of great significance in the medical domain. Hence, we evaluate the models' capability to synthesize PET scans that improve AD diagnostic accuracy. 

We conducted the two classification experiments as follows: (a) A 5-fold cross-validation experiment using the Paired dataset where 4 folds were used to train both the generation and classification networks followed by testing on the remaining fold. (b) The generation models were used as data augmentation tools by training them on the Paired dataset. The trained models were used to complete the Incomplete dataset by synthesizing the missing PET scans. AD classification was then conducted on the completed dataset.  

Classification results across all metrics for these two experiments are shown in Tables  \ref{table:class402}  and \ref{table:class581}, respectively. The results clearly indicate that the proposed GLA-GAN provides better AD classification accuracy than the competing methods.
We could not perform a direct comparison with the state-of-the-art classification due to differing experimental settings, particularly in terms of paired and completed data availability. These variations in dataset preparation and evaluation protocols make it challenging to ensure a fair and meaningful comparison.

    
\begin{table}[h!]
\renewcommand{\arraystretch}{1.6}
\resizebox{\columnwidth}{!}{\begin{tabular}{|c|c|c|c|c|c|c|}
\hline
\textbf{Method}                                                          & \textbf{ACC} ($\uparrow$)             & \textbf{SENS}  ($\uparrow$)        & \textbf{SPEC} ($\uparrow$)          & \textbf{F1-Score} ($\uparrow$)             & \textbf{MCC} ($\uparrow$) & \textbf{AUC}   ($\uparrow$)           \\ \hline
3D U-Net \cite{MRI2PET_MICCAI}                                                          & 0.8325$\pm$0.020        & 0.7895$\pm$0.021          & 0.8664$\pm$0.016          & 0.7910$\pm$0.023          & 0.6559$\pm$0.015 & 0.9191$\pm$0.005                \\ \hline
C-GAN  \cite{conditional}                             &0.8406$\pm$0.005          & 0.7983$\pm$0.007          & 0.8717$\pm$0.009          & 0.8004$\pm$0.005          & 0.6735$\pm$0.010 & 0.9247$\pm$0.005            \\ \hline
Cycle GAN  \cite{mritopet}                                                     & 0.8326$\pm$0.020          & 0.8192$\pm$0.020          &0.8795$\pm$0.080$^\dag$            & \textbf{0.8369}$\pm$\textbf{0.03} & 0.6917 $\pm$ 0.027  & 0.9209$\pm$0.008     \\ \hline 
{Local GAN}                                                      & {0.8409$\pm$0.004}$^\dag$           & {0.8302$\pm$0.004}$^\dag$          & {0.8554$\pm$0.003}           & {0.8145$\pm$0.010} & {0.6920$\pm$0.015}$^\dag$    & 0.9260$\pm$0.008$^\dag$            \\ \hline 
{Global GAN}                                                       & {0.8333$\pm$0.005}         & {0.8185$\pm$0.009}         & {0.8404$\pm$0.006}          & {0.8040$\pm$0.001} & {0.6790$\pm$0.004}  & 0.9004$\pm$0.006            \\ \hline

{GLA-GAN}                & \textbf{0.8456} $\pm$ \textbf{0.007}  & \textbf{0.8330} $\pm$ \textbf{0.006} & \textbf{0.8900}$\pm$\textbf{0.005} & 0.8256$\pm$ 0.007  & \textbf{0.7198} $\pm$ \textbf{0.013} & \textbf{0.9314}$\pm$\textbf{0.004}          \\ \hline
\end{tabular}}
\caption{AD Classification Results: Comparison of models across standard classification performance metrics for the Paired dataset. Results are shown as mean $\pm$ standard deviation. (Bold and $^\dag$ represent the first and second best, respectively)}
\label{table:class402}
\end{table}
\vspace*{-23pt}

\begin{table}[h!]
\renewcommand{\arraystretch}{1.4}
\resizebox{\linewidth}{!}{\begin{tabular}{|c|c|c|c|c|c|c|c|}
\hline
\textbf{Method}                                                          & \textbf{ACC} ($\uparrow$)             & \textbf{SENS}  ($\uparrow$)        & \textbf{SPEC} ($\uparrow$)          & \textbf{F1-Score} ($\uparrow$)             & \textbf{MCC} ($\uparrow$) & \textbf{AUC}   ($\uparrow$)           \\ \hline
Only MRI                                                           & 0.8293$\pm$0.016         & 0.803$\pm$0.035          & 0.8537$\pm$0.041        & 0.8047$\pm$0.022          & 0.6557$\pm$0.034  & 0.9035$\pm$0.019                \\ \hline
3D U-Net  \cite{MRI2PET_MICCAI}                                                         & 0.8289$\pm$0.007          & 0.7947$\pm$0.012          & 0.8597$\pm$0.003          & 0.8029$\pm$0.009          & 0.6559$\pm$0.015             & 0.9128$\pm$0.005 \\ \hline
C-GAN   \cite{conditional}                              & 0.8223$\pm$0.003          & 0.7819$\pm$0.005          & 0.8573$\pm$0.003          & 0.7940$\pm$0.003          & 0.6406$\pm$0.006    & 0.9072$\pm$0.001             \\ \hline
Cycle GAN \cite{mritopet}                                                      & 0.8352$\pm$0.020 $^\dag$         & 0.8133$\pm$0.018  $^\dag$        & 0.8796$\pm$0.040           & \textbf{0.8379}$\pm$\textbf{0.021} & 0.6949 $\pm$ 0.024 $^\dag$       & 0.9160$\pm$0.060 $^\dag$              \\ \hline 
{Local GAN}                                                      & {0.8287$\pm$0.004}          & {0.7690$\pm$0.010}         & {0.8907$\pm$0.001} $^\dag$          & {0.8119$\pm$0.003} & {0.6569$\pm$0.002}       & 0.9130$\pm$0.002             \\ \hline 
{Global GAN}                                                      & {0.8204$\pm$0.002}          & {0.7607$\pm$0.008}         & {0.8859$\pm$0.001}          & {0.7980$\pm$0.003} & {0.6420$\pm$0.010}          & 0.9021$\pm$0.001        \\ \hline 

{GLA-GAN}   & \textbf{0.8450} $\pm$ \textbf{0.004}  & \textbf{0.8343} $\pm$ \textbf{0.005} & \textbf{0.8919} $\pm$ \textbf{0.004} & 0.8292$\pm$ 0.010  & \textbf{0.7235} $\pm$ \textbf{0.001} & \textbf{0.9299}$\pm$\textbf{0.003}                 \\ \hline 
\end{tabular}}

\caption{AD Classification Results: Comparison of models across standard classification performance metrics for the Complete datasets (Incomplete dataset augmented with synthesized versions of missing PET scans from each model). Results are shown as mean $\pm$ standard deviation. (Bold and $^\dag$ represent the first and second best, respectively)}
\label{table:class581}
\end{table}

\subsection{Ablation Studies}
\begin{figure}[h!]
  \centering
 \includegraphics[width=0.9\textwidth]{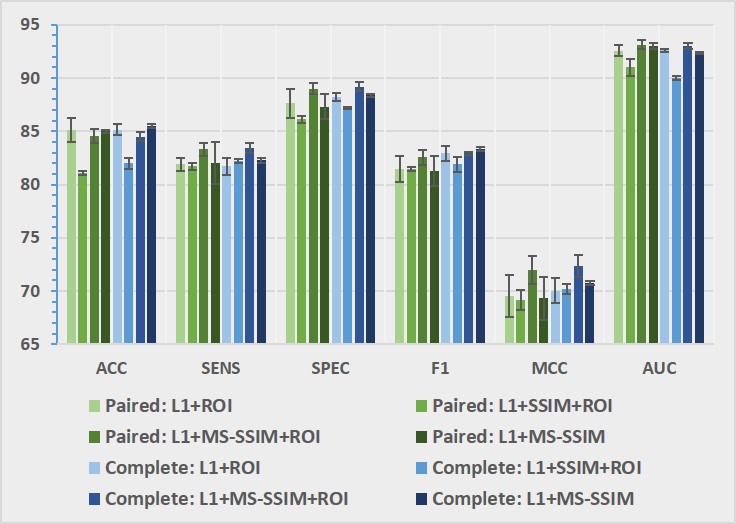}
  \caption{Evaluation of the GLA-GAN model with different combinations of loss components across standard classification performance metrics on both paired and complete datasets.}
   \label{fig:ablation1}
\end{figure}

To understand the importance of different loss components and their contribution to the synthesis task, we compare four variations of GLA-GAN models with different combinations of penalty terms: $\mathcal{L}_{L1}, \  \mathcal{L}_{\mathrm{SSIM}}, \  \mathcal{L}_{\mathrm{MS-SSIM}}, \ $ and 
$\mathcal{L}_{ROI}$.  These variants of GLA-GAN were applied to both the paired and complete datasets. The results depicted in Figure \ref{fig:ablation1} establish that ROI-based loss contributes the most to improving AD classification. Although classification accuracy is relatively low for ROI combined with SSIM loss, other metrics had no significant differences. It can also be noticed that replacing SSIM with MS-SSIM loss improves the results, indicating the necessity to ensure structural integrity at multiple scales. These results suggest that the constraints imposed by the $\mathcal{L}_{MS-SSIM}$ and $\mathcal{L}_{ROI}$ assist the hybrid model to not only capture the overall structure but also preserve the local, regional details. These enhancements to the generation quality, in turn, lead to the improved multi-modal classification of AD using synthesized PET scans.

\begin{figure}[!htbp]
  \centering
 \includegraphics[width=0.9\textwidth]{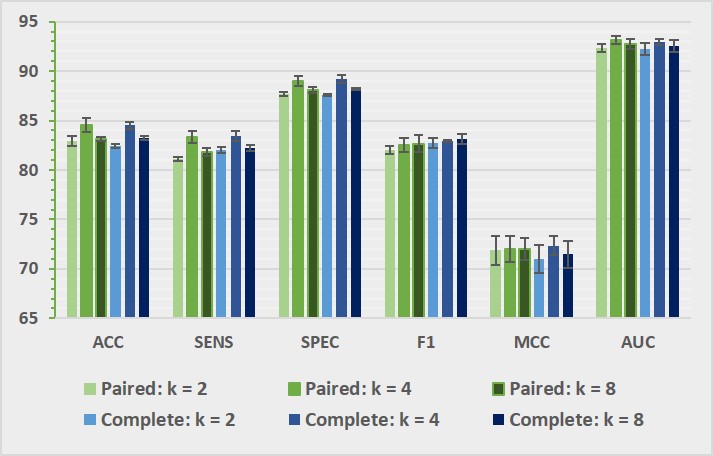}
  \caption{Evaluation of the GLA-GAN model with a varying number of patches ($K$) used in the local module across standard classification performance metrics on both paired dataset and complete dataset.}
   \label{fig:ablation2}
\end{figure}

We have already established the importance of having both global and local contextual modules to generate superior-quality synthesized PET scans. Subsequently, we further explored the impact of the number of non-overlapping patches used by the local module on the model's performance. The results of the experiment conducted with $K = 2, 4,$ and $8$ are presented in Figure \ref{fig:ablation2}. Contrary to our expectations, the performance did not increase with increasing $K$ but with $K=4$ providing the best results. However, further investigation is needed to explore if anatomically informed rather than generic cuboid patches would be more efficient.

\subsection{Understanding GLA-GAN}
Recently, the explainability of deep learning models has attracted increasing attention. Here, we attempt to gain insights into their inner workings using interpretation strategies such as exploratory analysis and visualization.

\subsubsection{Visualization} 
Visualization of saliency maps has become a standard tool for interpreting the attention of neural networks based on the backpropagation of gradients. These probabilistic maps are only effective when specific objects or targets of interest exist. For example, faces in computer vision applications or lesions in the medical domain. However, for a disorder like AD that is characterized by gross brain atrophy involving many regions and varying significantly across subjects and through disease progression, saliency maps are not very informative.

\begin{figure*}[!t]
  \centering
  \includegraphics[width=0.98\textwidth]{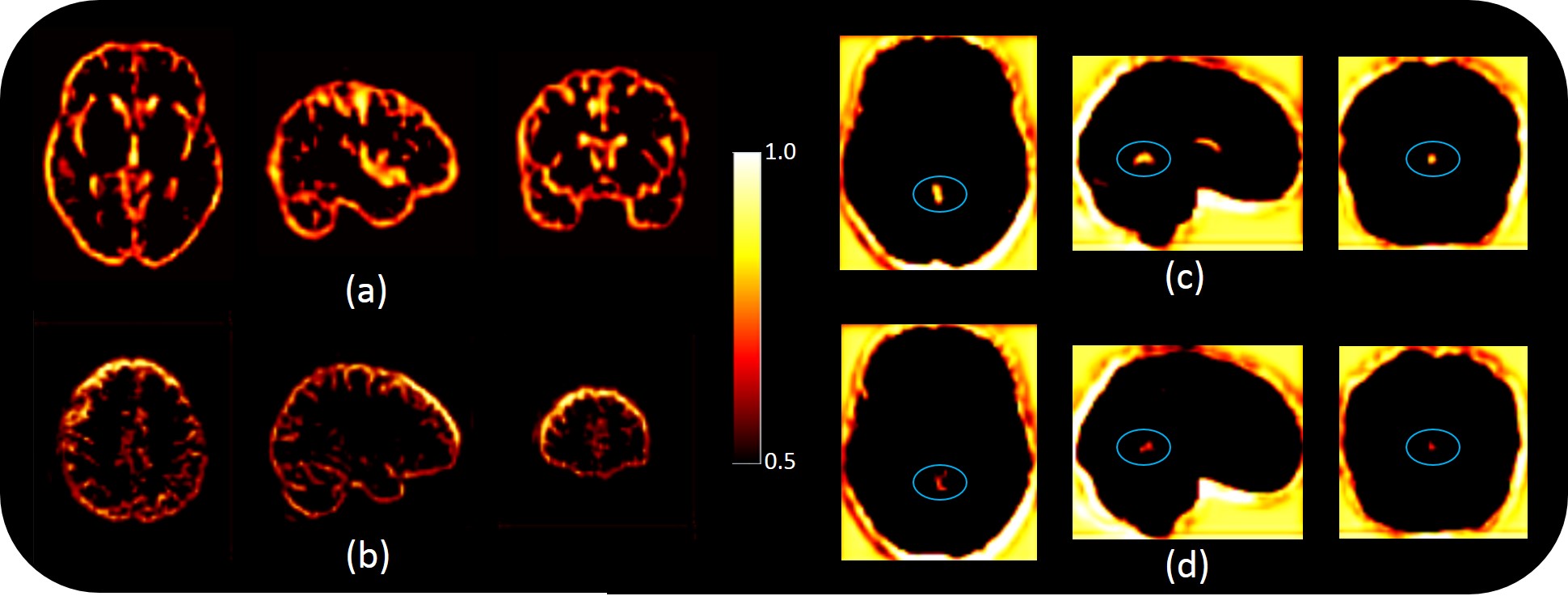}
  \caption{Visualization of internal features from the global module:  (a) Feature that detects the overall uptake pattern covering regions most commonly associated with AD (cortex, subcortex, ventricles, etc.) (b) Feature that perceives relatively significant importance of superior frontal gyrus in AD  (c) \& (d) Feature that detects the vermis region of the cerebellum (associated with early stages of AD) validated by intensity differences between a CN and an AD sample respectively.}
   \label{fig:GANVis}
\end{figure*}

Although many of the features did not lend themselves readily to human interpretability, we were able to identify some features from the global module ($G_g$) of the GLA-GAN that were relevant to this particular task. Some commonly used biomarkers for Alzheimer’s include cortical and subcortical atrophy and dilation of ventricles. One of the features characterized by these specific brain regions is shown in Figure \ref{fig:GANVis} (a).  Furthermore, several neuroimaging studies have identified specific brain regions most prominently affected by the disease \cite{Sawyer217}. The feature depicted in Figure \ref{fig:GANVis} (b) highlights the relative significance of the superior frontal gyrus region. Metabolic changes in this particular region have been associated with cognitive decline \cite{Hernandez2018}. In addition to these overall features, we identified a feature that emphasized the potential contribution of a specific brain region for AD diagnosis. Figures \ref{fig:GANVis} (c) and (d) display the same feature that illustrates the importance of the vermis region of the cerebellum with intensity differences between the normal and AD subjects, respectively. This particular region has been implicated in being affected in the early stages of the disease \cite{jacobs2018cerebellum}, which could be crucial for timely diagnosis.

It was very encouraging to detect internal features from the global module that exhibited strong connotations characteristic of Alzheimer’s disease.  Unfortunately, as the individual subnetworks of the local generator module are trained independently, the concatenated versions of the local features were not easy to decipher. Subsequently, we found it difficult to identify feature maps from the end-layer that could be interpreted in terms of contributions associated with global and local components.

\subsubsection{Interpolation} 

\begin{figure*}[!htbp]
  \centering
  \includegraphics[width=0.95\textwidth]{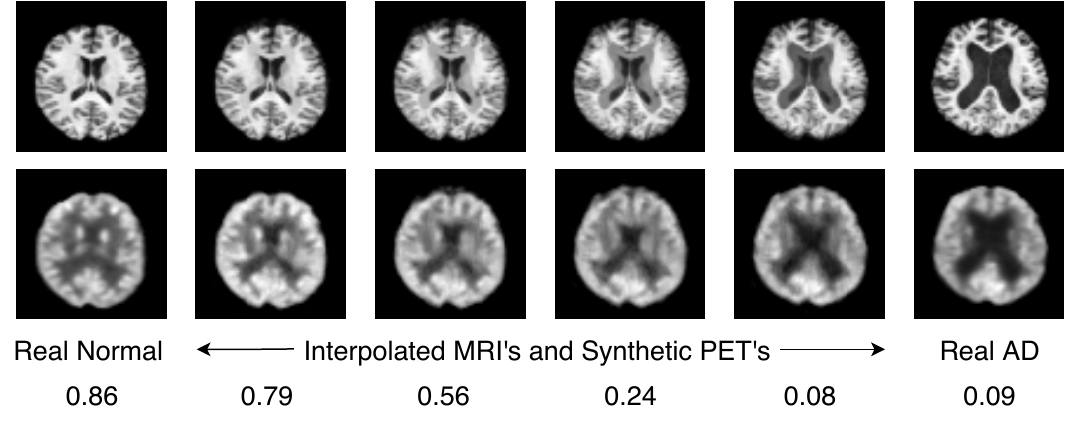}
  \caption{MRI interpolations between one Normal and AD pair. Left-most and right columns show pairs of real MRI and PET scans. Intermediate columns depict linear interpolations in the MRI space and their corresponding generated PET scans. The last row of numbers indicates the classification probability as Normal.   }
   \label{fig:interp}
\end{figure*}

 GANs are prone to overfitting, more so with medical images that have high dimensionality and low sample size. To ensure that our model is not memorizing training images, we performed linear interpolation in MRI space between two MR images chosen from different classes to clearly see the effect of the transition from one class to the other. Then, PET scans corresponding to interpolated MR images are generated using our model. If the model is overfitting, we expect to see discrete transitions in the interpolated images \cite{conditional} As Figure \ref{fig:interp} shows, the generated PET scans follow a smooth transition with no artifacts indicating that the model has learned the underlying distribution well. Moreover, we classified the generated scans from interpolation, and as expected, the probability of being classified as normal also changes gradually from high to low as we approach AD.

\subsection{Limitations}

Our work also has some limitations which need to be acknowledged. Firstly, introducing local sub-networks to capture finer texture details leads to an increase in the number of network parameters compared to the competing global methods. We believe that using more semantically segmented patches and exploiting sparsity constraints in the future can lead to more optimized networks with similar performance.  

Secondly, the focus of this study has been limited to discriminating AD from CN. However, mild cognitive impairment (MCI) is considered an intermediate state between normal aging and the onset of AD. Further validation is needed to assess the effectiveness of the proposed GLA-GAN for a 3-way classification (AD-MCI-CN), which is crucial for early diagnosis. 

Lastly, an exhaustive search was not conducted to find all or the most meaningful internal GAN units.  Due to the high dimensionality of the images being handled and the large number of internal units, identifying meaningful features is a very tedious and time-consuming task. Future efforts can be directed towards more systematic approaches that rely on the agreement between activation units and anatomically segmented synthetic scans. Along the lines of  \cite{GANDissect}, it would be very interesting to explore the causal effect of interpretable units in explaining Alzheimer's disease progression.

\section{Conclusion}
This paper presents a globally and locally aware GAN framework for MRI-to-PET cross-modality transfer. The proposed multi-path GAN architecture aids in concurrently capturing both global structures as well as local textures to enhance the quality of synthesized PET scans. Furthermore, the standard adversarial loss in GANs is supplemented with $L_1$ loss to ensure voxel level fidelity of intensity values, MS-SSIM loss to ensure structural consistency at multiple scales, and ROI loss to enforce local contextual integrity to discern variation in regional sensitivity to AD. The overall framework and the combined synthesis objective function were designed to help the generative model reliably learn the underlying bimodal data distribution. Experimental results demonstrate that our model generates PET scans with better image quality and enriched features that improve AD diagnostic accuracy compared to state-of-the-art methods. Finally, our attempt to visualize intermediate feature maps exhibits the aptitude of GLA-GAN to automatically detect both common and differential FDG uptake patterns in PET images.

\section*{Declarations}
\begin{itemize}
\item \textbf{Conflict of Interest} - The authors declare that they have no conflict of interest.
\item \textbf{Funding} - The authors did not receive support from any organization for the submitted work.

\item \textbf{Author contribution} - Conceptualization: Apoorva Sikka, Skand Peri, Deepti R. Bathula; Methodology: Apoorva Sikka, Skand Peri; Formal analysis and investigation: Apoorva Sikka, Skand Peri, Jitender Singh Virk, Usma Niyaz; Writing - original draft preparation: Apoorva Sikka; Writing - review and editing: Usma Niyaz, Deepti R. Bathula
    \item \textbf{Data Availability Statement} - This research used open-access public datasets for research investigation.
    \item \textbf{Research Involving Human and /or Animals} - Not applicable
    \item \textbf{Informed consent} - Not applicable
    
\end{itemize}

\bibliography{sn}


\begin{thebibliography}{52}
\ifx \bisbn   \undefined \def \bisbn  #1{ISBN #1}\fi
\ifx \binits  \undefined \def \binits#1{#1}\fi
\ifx \bauthor  \undefined \def \bauthor#1{#1}\fi
\ifx \batitle  \undefined \def \batitle#1{#1}\fi
\ifx \bjtitle  \undefined \def \bjtitle#1{#1}\fi
\ifx \bvolume  \undefined \def \bvolume#1{\textbf{#1}}\fi
\ifx \byear  \undefined \def \byear#1{#1}\fi
\ifx \bissue  \undefined \def \bissue#1{#1}\fi
\ifx \bfpage  \undefined \def \bfpage#1{#1}\fi
\ifx \blpage  \undefined \def \blpage #1{#1}\fi
\ifx \burl  \undefined \def \burl#1{\textsf{#1}}\fi
\ifx \doiurl  \undefined \def \doiurl#1{\url{https://doi.org/#1}}\fi
\ifx \betal  \undefined \def \betal{\textit{et al.}}\fi
\ifx \binstitute  \undefined \def \binstitute#1{#1}\fi
\ifx \binstitutionaled  \undefined \def \binstitutionaled#1{#1}\fi
\ifx \bctitle  \undefined \def \bctitle#1{#1}\fi
\ifx \beditor  \undefined \def \beditor#1{#1}\fi
\ifx \bpublisher  \undefined \def \bpublisher#1{#1}\fi
\ifx \bbtitle  \undefined \def \bbtitle#1{#1}\fi
\ifx \bedition  \undefined \def \bedition#1{#1}\fi
\ifx \bseriesno  \undefined \def \bseriesno#1{#1}\fi
\ifx \blocation  \undefined \def \blocation#1{#1}\fi
\ifx \bsertitle  \undefined \def \bsertitle#1{#1}\fi
\ifx \bsnm \undefined \def \bsnm#1{#1}\fi
\ifx \bsuffix \undefined \def \bsuffix#1{#1}\fi
\ifx \bparticle \undefined \def \bparticle#1{#1}\fi
\ifx \barticle \undefined \def \barticle#1{#1}\fi
\bibcommenthead
\ifx \bconfdate \undefined \def \bconfdate #1{#1}\fi
\ifx \botherref \undefined \def \botherref #1{#1}\fi
\ifx \url \undefined \def \url#1{\textsf{#1}}\fi
\ifx \bchapter \undefined \def \bchapter#1{#1}\fi
\ifx \bbook \undefined \def \bbook#1{#1}\fi
\ifx \bcomment \undefined \def \bcomment#1{#1}\fi
\ifx \oauthor \undefined \def \oauthor#1{#1}\fi
\ifx \citeauthoryear \undefined \def \citeauthoryear#1{#1}\fi
\ifx \endbibitem  \undefined \def \endbibitem {}\fi
\ifx \bconflocation  \undefined \def \bconflocation#1{#1}\fi
\ifx \arxivurl  \undefined \def \arxivurl#1{\textsf{#1}}\fi
\csname PreBibitemsHook\endcsname

\bibitem[\protect\citeauthoryear{Lorenzi et~al.}{2016}]{multimodal}
\begin{barticle}
\bauthor{\bsnm{Lorenzi}, \binits{M.}},
\bauthor{\bsnm{Simpson}, \binits{I.}},
\bauthor{\bsnm{Mendelson}, \binits{A.}},
\bauthor{\bsnm{Vos}, \binits{S.}},
\bauthor{\bsnm{Cardoso}, \binits{M.J.}},
\bauthor{\bsnm{Modat}, \binits{M.}},
\bauthor{\bsnm{Schott}, \binits{J.}},
\bauthor{\bsnm{Ourselin}, \binits{S.}}:
\batitle{Multimodal image analysis in alzheimer's disease via statistical modelling of non-local intensity correlations}.
\bjtitle{Scientific reports}
\bvolume{6},
\bfpage{22161}
(\byear{2016})
\doiurl{10.1038/srep22161}
\end{barticle}
\endbibitem

\bibitem[\protect\citeauthoryear{Gray et~al.}{2011}]{fdgAD}
\begin{bchapter}
\bauthor{\bsnm{Gray}, \binits{K.}},
\bauthor{\bsnm{Wolz}, \binits{R.}},
\bauthor{\bsnm{Keihaninejad}, \binits{S.}},
\bauthor{\bsnm{Heckemann}, \binits{R.}},
\bauthor{\bsnm{Aljabar}, \binits{P.}},
\bauthor{\bsnm{Hammers}, \binits{A.}},
\bauthor{\bsnm{Rueckert}, \binits{D.}}:
\bctitle{Regional analysis of fdg-pet for use in the classification of alzheimer's disease}.
In: \bbtitle{IEEE International Symposium on Biomedical Imaging(ISBI)},
pp. \bfpage{1082}--\blpage{1085}
(\byear{2011}).
\doiurl{10.1109/ISBI.2011.5872589}
\end{bchapter}
\endbibitem

\bibitem[\protect\citeauthoryear{Johnson et~al.}{2012}]{johnson2012brain}
\begin{barticle}
\bauthor{\bsnm{Johnson}, \binits{K.}},
\bauthor{\bsnm{Fox}, \binits{N.}},
\bauthor{\bsnm{Sperling}, \binits{R.}},
\bauthor{\bsnm{Klunk}, \binits{W.}}:
\batitle{Brain imaging in alzheimer disease}.
\bjtitle{Cold Spring Harbor perspectives in medicine}
\bvolume{2},
\bfpage{006213}
(\byear{2012})
\doiurl{10.1101/cshperspect.a006213}
\end{barticle}
\endbibitem

\bibitem[\protect\citeauthoryear{Zhang et~al.}{2017}]{zhang2017pet}
\begin{barticle}
\bauthor{\bsnm{Zhang}, \binits{X.}},
\bauthor{\bsnm{Yang}, \binits{Z.}},
\bauthor{\bsnm{Lu}, \binits{G.}},
\bauthor{\bsnm{Yang}, \binits{G.}},
\bauthor{\bsnm{Zhang}, \binits{L.}}:
\batitle{Pet/mr imaging: New frontier in alzheimer's disease and other dementias}.
\bjtitle{Frontiers in Molecular Neuroscience}
\bvolume{10},
\bfpage{343}
(\byear{2017})
\doiurl{10.3389/fnmol.2017.00343}
\end{barticle}
\endbibitem

\bibitem[\protect\citeauthoryear{Cheng and Liu}{2017}]{CNNmulti}
\begin{bchapter}
\bauthor{\bsnm{Cheng}, \binits{D.}},
\bauthor{\bsnm{Liu}, \binits{M.}}:
\bctitle{Cnns based multi-modality classification for ad diagnosis}.
In: \bbtitle{10th International Congress on Image and Signal Processing, BioMedical Engineering and Informatics (CISP-BMEI)},
pp. \bfpage{1}--\blpage{5}
(\byear{2017}).
\doiurl{10.1109/CISP-BMEI.2017.8302281}
\end{bchapter}
\endbibitem

\bibitem[\protect\citeauthoryear{Huynh et~al.}{2015}]{huynh2015estimating}
\begin{barticle}
\bauthor{\bsnm{Huynh}, \binits{T.}},
\bauthor{\bsnm{Gao}, \binits{Y.}},
\bauthor{\bsnm{Kang}, \binits{J.}},
\bauthor{\bsnm{Wang}, \binits{L.}},
\bauthor{\bsnm{Zhang}, \binits{P.}},
\bauthor{\bsnm{Lian}, \binits{J.}},
\bauthor{\bsnm{Shen}, \binits{D.}}:
\batitle{Estimating ct image from mri data using structured random forest and auto-context model}.
\bjtitle{IEEE transactions on medical imaging}
\bvolume{35},
\bfpage{174}--\blpage{183}
(\byear{2015})
\doiurl{10.1109/TMI.2015.2461533}
\end{barticle}
\endbibitem

\bibitem[\protect\citeauthoryear{Kang et~al.}{2015}]{lowdose}
\begin{barticle}
\bauthor{\bsnm{Kang}, \binits{J.}},
\bauthor{\bsnm{Gao}, \binits{Y.}},
\bauthor{\bsnm{Shi}, \binits{F.}},
\bauthor{\bsnm{Lin}, \binits{W.}},
\bauthor{\bsnm{Shen}, \binits{D.}}:
\batitle{Prediction of standard-dose brain pet image by using mri and low-dose brain [(18)f]fdg pet images}.
\bjtitle{Medical physics}
\bvolume{42},
\bfpage{5301}
(\byear{2015})
\doiurl{10.1118/1.4928400}
\end{barticle}
\endbibitem

\bibitem[\protect\citeauthoryear{Jog et~al.}{2014}]{jog2014random}
\begin{barticle}
\bauthor{\bsnm{Jog}, \binits{A.}},
\bauthor{\bsnm{Carass}, \binits{A.}},
\bauthor{\bsnm{Pham}, \binits{D.}},
\bauthor{\bsnm{Prince}, \binits{J.}}:
\batitle{Random forest flair reconstruction from t1, t2, and pd-weighted mri}.
\bjtitle{11th International Symposium on Biomedical Imaging (ISBI)}
\bvolume{2014},
\bfpage{1079}--\blpage{1082}
(\byear{2014})
\doiurl{10.1109/ISBI.2014.6868061}
\end{barticle}
\endbibitem

\bibitem[\protect\citeauthoryear{Goodfellow et~al.}{2014}]{GAN}
\begin{barticle}
\bauthor{\bsnm{Goodfellow}, \binits{I.}},
\bauthor{\bsnm{Pouget-Abadie}, \binits{J.}},
\bauthor{\bsnm{Mirza}, \binits{M.}},
\bauthor{\bsnm{Xu}, \binits{B.}},
\bauthor{\bsnm{Warde-Farley}, \binits{D.}},
\bauthor{\bsnm{Ozair}, \binits{S.}},
\bauthor{\bsnm{Courville}, \binits{A.}},
\bauthor{\bsnm{Bengio}, \binits{Y.}}:
\batitle{Generative adversarial networks}.
\bjtitle{Advances in Neural Information Processing Systems}
\bvolume{3},
\bfpage{2672}--\blpage{2680}
(\byear{2014})
\doiurl{10.1145/3422622}
\end{barticle}
\endbibitem

\bibitem[\protect\citeauthoryear{Radford et~al.}{2016}]{DCGAN_ICLR}
\begin{bchapter}
\bauthor{\bsnm{Radford}, \binits{A.}},
\bauthor{\bsnm{Metz}, \binits{L.}},
\bauthor{\bsnm{Chintala}, \binits{S.}}:
\bctitle{Unsupervised representation learning with deep convolutional generative adversarial networks}.
In: \bbtitle{4th International Conference on Learning Representations (ICLR)},
vol. \bseriesno{abs/1511.06434}
(\byear{2016}).
\doiurl{10.1109/AIAR.2018.8769811}
\end{bchapter}
\endbibitem

\bibitem[\protect\citeauthoryear{Zhang et~al.}{2017}]{stackgan}
\begin{bchapter}
\bauthor{\bsnm{Zhang}, \binits{H.}},
\bauthor{\bsnm{Xu}, \binits{T.}},
\bauthor{\bsnm{Li}, \binits{H.}}:
\bctitle{Stackgan: Text to photo-realistic image synthesis with stacked generative adversarial networks}.
In: \bbtitle{International Conference on Computer Vision (ICCV)},
pp. \bfpage{5908}--\blpage{5916}
(\byear{2017}).
\doiurl{10.1109/ICCV.2017.629}
\end{bchapter}
\endbibitem

\bibitem[\protect\citeauthoryear{Yang et~al.}{2017}]{denoising}
\begin{barticle}
\bauthor{\bsnm{Yang}, \binits{Q.}},
\bauthor{\bsnm{Yan}, \binits{P.}},
\bauthor{\bsnm{Zhang}, \binits{Y.}},
\bauthor{\bsnm{Yu}, \binits{H.}},
\bauthor{\bsnm{Shi}, \binits{Y.}},
\bauthor{\bsnm{Mou}, \binits{X.}},
\bauthor{\bsnm{Kalra}, \binits{M.}},
\bauthor{\bsnm{Zhang}, \binits{Y.}},
\bauthor{\bsnm{Sun}, \binits{L.}},
\bauthor{\bsnm{Wang}, \binits{G.}}:
\batitle{Low-dose ct image denoising using a generative adversarial network with wasserstein distance and perceptual loss}.
\bjtitle{IEEE Transactions on Medical Imaging}
\bvolume{37},
\bfpage{1348}--\blpage{1357}
(\byear{2017})
\doiurl{10.1109/TMI.2018.2827462}
\end{barticle}
\endbibitem

\bibitem[\protect\citeauthoryear{Wolterink et~al.}{2017}]{mrtoct}
\begin{bchapter}
\bauthor{\bsnm{Wolterink}, \binits{J.}},
\bauthor{\bsnm{Dinkla}, \binits{A.}},
\bauthor{\bsnm{Savenije}, \binits{M.}},
\bauthor{\bsnm{Seevinck}, \binits{P.}},
\bauthor{\bsnm{Berg}, \binits{C.}},
\bauthor{\bsnm{Išgum}, \binits{I.}}:
\bctitle{Deep mr to ct synthesis using unpaired data}.
In: \bbtitle{MICCAI Workshop on Simulation and Synthesis in Medical Imaging},
vol. \bseriesno{10557},
pp. \bfpage{14}--\blpage{23}
(\byear{2017}).
\doiurl{10.1007/978-3-319-68127-6_2}
\end{bchapter}
\endbibitem

\bibitem[\protect\citeauthoryear{Frid-Adar et~al.}{2018}]{frid2018gan}
\begin{barticle}
\bauthor{\bsnm{Frid-Adar}, \binits{M.}},
\bauthor{\bsnm{Diamant}, \binits{I.}},
\bauthor{\bsnm{Klang}, \binits{E.}},
\bauthor{\bsnm{Amitai}, \binits{M.}},
\bauthor{\bsnm{Goldberger}, \binits{J.}},
\bauthor{\bsnm{Greenspan}, \binits{H.}}:
\batitle{Gan-based synthetic medical image augmentation for increased cnn performance in liver lesion classification}.
\bjtitle{Neurocomputing}
\bvolume{321},
\bfpage{321}--\blpage{331}
(\byear{2018})
\doiurl{10.1016/j.neucom.2018.09.013}
\end{barticle}
\endbibitem

\bibitem[\protect\citeauthoryear{Han et~al.}{2018a}]{han2018gan}
\begin{bchapter}
\bauthor{\bsnm{Han}, \binits{C.}},
\bauthor{\bsnm{Hayashi}, \binits{H.}},
\bauthor{\bsnm{Rundo}, \binits{L.}},
\bauthor{\bsnm{Araki}, \binits{R.}},
\bauthor{\bsnm{Shimoda}, \binits{W.}},
\bauthor{\bsnm{Muramatsu}, \binits{S.}},
\bauthor{\bsnm{Furukawa}, \binits{Y.}},
\bauthor{\bsnm{Mauri}, \binits{G.}},
\bauthor{\bsnm{Nakayama}, \binits{H.}}:
\bctitle{Gan-based synthetic brain mr image generation}.
In: \bbtitle{15th International Symposium on Biomedical Imaging (ISBI)},
pp. \bfpage{734}--\blpage{738}
(\byear{2018}).
\doiurl{10.1109/ISBI.2018.8363678}
\end{bchapter}
\endbibitem

\bibitem[\protect\citeauthoryear{Han et~al.}{2018b}]{han2018spine}
\begin{barticle}
\bauthor{\bsnm{Han}, \binits{Z.}},
\bauthor{\bsnm{Wei}, \binits{B.}},
\bauthor{\bsnm{Mercado}, \binits{A.}},
\bauthor{\bsnm{Leung}, \binits{S.}},
\bauthor{\bsnm{Li}, \binits{S.}}:
\batitle{Spine-gan: Semantic segmentation of multiple spinal structures}.
\bjtitle{Medical Image Analysis}
\bvolume{50},
\bfpage{23}--\blpage{35}
(\byear{2018})
\doiurl{10.1016/j.media.2018.08.005}
\end{barticle}
\endbibitem

\bibitem[\protect\citeauthoryear{Sevetlidis et~al.}{2016}]{sevetlidis2016whole}
\begin{bchapter}
\bauthor{\bsnm{Sevetlidis}, \binits{V.}},
\bauthor{\bsnm{Giuffrida}, \binits{V.}},
\bauthor{\bsnm{Tsaftaris}, \binits{S.}}:
\bctitle{Whole image synthesis using a deep encoder-decoder network}.
In: \bbtitle{MICCAI Workshop on Simulation and Synthesis in Medical Imaging},
vol. \bseriesno{9968},
pp. \bfpage{127}--\blpage{137}
(\byear{2016}).
\doiurl{10.1007/978-3-319-46630-9_13}
\end{bchapter}
\endbibitem

\bibitem[\protect\citeauthoryear{Kingma and Welling}{2014}]{kingma2013auto}
\begin{bchapter}
\bauthor{\bsnm{Kingma}, \binits{D.}},
\bauthor{\bsnm{Welling}, \binits{M.}}:
\bctitle{Auto-encoding variational bayes}.
In: \bbtitle{2nd International Conference on Learning Representations (ICLR)}
(\byear{2014}).
\doiurl{10.48550/arXiv.1312.6114}
\end{bchapter}
\endbibitem

\bibitem[\protect\citeauthoryear{Nie et~al.}{2017}]{ACM-GAN}
\begin{bchapter}
\bauthor{\bsnm{Nie}, \binits{D.}},
\bauthor{\bsnm{Trullo}, \binits{R.}},
\bauthor{\bsnm{Lian}, \binits{J.}},
\bauthor{\bsnm{Petitjean}, \binits{C.}},
\bauthor{\bsnm{Ruan}, \binits{S.}},
\bauthor{\bsnm{Wang}, \binits{Q.}}:
\bctitle{Medical image synthesis with context-aware generative adversarial networks}.
In: \bbtitle{Medical Image Computing and Computer Assisted Intervention (MICCAI)},
vol. \bseriesno{10435},
pp. \bfpage{417}--\blpage{425}
(\byear{2017}).
\doiurl{10.1007/978-3-319-66179-7_48}
\end{bchapter}
\endbibitem

\bibitem[\protect\citeauthoryear{Jin et~al.}{2019}]{Jin2019DeepCT}
\begin{barticle}
\bauthor{\bsnm{Jin}, \binits{C.}},
\bauthor{\bsnm{Kim}, \binits{H.}},
\bauthor{\bsnm{Liu}, \binits{M.}},
\bauthor{\bsnm{Jung}, \binits{W.}},
\bauthor{\bsnm{Joo}, \binits{S.}},
\bauthor{\bsnm{Park}, \binits{E.}},
\bauthor{\bsnm{Ahn}, \binits{Y.}},
\bauthor{\bsnm{Han}, \binits{I.}},
\bauthor{\bsnm{Lee}, \binits{J.}},
\bauthor{\bsnm{Cui}, \binits{X.}}:
\batitle{Deep ct to mr synthesis using paired and unpaired data}.
\bjtitle{Sensors}
\bvolume{19},
\bfpage{2361}
(\byear{2019})
\doiurl{10.3390/s19102361}
\end{barticle}
\endbibitem

\bibitem[\protect\citeauthoryear{Ben-Cohen et~al.}{}]{cttopet}
\begin{botherref}
\oauthor{\bsnm{Ben-Cohen}, \binits{A.}},
\oauthor{\bsnm{Klang}, \binits{E.}},
\oauthor{\bsnm{Raskin}, \binits{S.}},
\oauthor{\bsnm{Soffer}, \binits{S.}},
\oauthor{\bsnm{Ben-Haim}, \binits{S.}},
\oauthor{\bsnm{Konen}, \binits{E.}},
\oauthor{\bsnm{Amitai}, \binits{M.}},
\oauthor{\bsnm{Greenspan}, \binits{H.}}:
Cross-modality synthesis from ct to pet using fcn and gan networks for improved automated lesion detection.
Engineering Applications of Artificial Intelligence
\textbf{78},
186--194
\doiurl{10.1016/j.engappai.2018.11.013}
\end{botherref}
\endbibitem

\bibitem[\protect\citeauthoryear{Wei et~al.}{2019}]{wei2019predicting}
\begin{barticle}
\bauthor{\bsnm{Wei}, \binits{W.}},
\bauthor{\bsnm{Poirion}, \binits{E.}},
\bauthor{\bsnm{Bodini}, \binits{B.}},
\bauthor{\bsnm{Durrleman}, \binits{S.}},
\bauthor{\bsnm{Ayache}, \binits{N.}},
\bauthor{\bsnm{Stankoff}, \binits{B.}},
\bauthor{\bsnm{Colliot}, \binits{O.}}:
\batitle{Predicting pet-derived demyelination from multimodal mri using sketcher-refiner adversarial training for multiple sclerosis}.
\bjtitle{Medical Image Analysis}
\bvolume{58},
\bfpage{101546}
(\byear{2019})
\doiurl{10.1016/j.media.2019.101546}
\end{barticle}
\endbibitem

\bibitem[\protect\citeauthoryear{li et~al.}{2014}]{patch}
\begin{bchapter}
\bauthor{\bsnm{li}, \binits{R.}},
\bauthor{\bsnm{Zhang}, \binits{W.}},
\bauthor{\bsnm{Suk}, \binits{H.-I.}},
\bauthor{\bsnm{Wang}, \binits{L.}},
\bauthor{\bsnm{Li}, \binits{J.}},
\bauthor{\bsnm{Shen}, \binits{D.}},
\bauthor{\bsnm{Ji}, \binits{S.}}:
\bctitle{Deep learning based imaging data completion for improved brain disease diagnosis}.
In: \bbtitle{Medical Image Computing and Computer Assisted Intervention(MICCAI)},
vol. \bseriesno{17},
pp. \bfpage{305}--\blpage{12}
(\byear{2014}).
\doiurl{10.1007/978-3-319-10443-0_39}
\end{bchapter}
\endbibitem

\bibitem[\protect\citeauthoryear{Sikka et~al.}{2018}]{MRI2PET_MICCAI}
\begin{bchapter}
\bauthor{\bsnm{Sikka}, \binits{A.}},
\bauthor{\bsnm{Peri}, \binits{S.V.}},
\bauthor{\bsnm{Bathula}, \binits{D.R.}}:
\bctitle{Mri to fdg-pet: Cross-modal synthesis using 3d u-net for multi-modal alzheimer's classification}.
In: \bbtitle{MICCAI Workshop on Simulation and Synthesis in Medical Imaging},
vol. \bseriesno{11037},
pp. \bfpage{80}--\blpage{89}
(\byear{2018}).
\doiurl{10.1007/978-3-030-00536-8_9}
\end{bchapter}
\endbibitem

\bibitem[\protect\citeauthoryear{Pan et~al.}{2018}]{mritopet}
\begin{bchapter}
\bauthor{\bsnm{Pan}, \binits{Y.}},
\bauthor{\bsnm{Liu}, \binits{M.}},
\bauthor{\bsnm{Lian}, \binits{C.}},
\bauthor{\bsnm{Zhou}, \binits{T.}},
\bauthor{\bsnm{Xia}, \binits{Y.}}:
\bctitle{Synthesizing missing pet from mri with cycle-consistent generative adversarial networks for alzheimer’s disease diagnosis}.
In: \bbtitle{Medical Image Computing and Computer Assisted Intervention (MICCAI)},
vol. \bseriesno{11072},
pp. \bfpage{455}--\blpage{463}
(\byear{2018}).
\doiurl{10.1007/978-3-030-00931-1_52}
\end{bchapter}
\endbibitem

\bibitem[\protect\citeauthoryear{Sun et~al.}{2019}]{Glow}
\begin{bchapter}
\bauthor{\bsnm{Sun}, \binits{H.}},
\bauthor{\bsnm{Mehta}, \binits{R.}},
\bauthor{\bsnm{Zhou}, \binits{H.}},
\bauthor{\bsnm{Huang}, \binits{Z.}},
\bauthor{\bsnm{Johnson}, \binits{S.}},
\bauthor{\bsnm{Prabhakaran}, \binits{V.}},
\bauthor{\bsnm{Singh}, \binits{V.}}:
\bctitle{Dual-glow: Conditional flow-based generative model for modality transfer}.
In: \bbtitle{International Conference on Computer Vision (ICCV)},
vol. \bseriesno{2019},
pp. \bfpage{10610}--\blpage{10619}
(\byear{2019}).
\doiurl{10.1109/ICCV.2019.01071}
\end{bchapter}
\endbibitem

\bibitem[\protect\citeauthoryear{Kingma and Dhariwal}{2018}]{kingma2018glow}
\begin{bchapter}
\bauthor{\bsnm{Kingma}, \binits{D.P.}},
\bauthor{\bsnm{Dhariwal}, \binits{P.}}:
\bctitle{Glow: Generative flow with invertible 1x1 convolutions}.
In: \bbtitle{Advances in Neural Information Processing Systems},
vol. \bseriesno{31},
pp. \bfpage{10215}--\blpage{10224}
(\byear{2018}).
\doiurl{10.48550/arXiv.1807.03039}
\end{bchapter}
\endbibitem

\bibitem[\protect\citeauthoryear{Johnson et~al.}{2016}]{perceptual}
\begin{bchapter}
\bauthor{\bsnm{Johnson}, \binits{J.}},
\bauthor{\bsnm{Alahi}, \binits{A.}},
\bauthor{\bsnm{Fei-Fei}, \binits{L.}}:
\bctitle{Perceptual losses for real-time style transfer and super-resolution}.
In: \bbtitle{European Conference on Computer Vision (ECCV)},
vol. \bseriesno{9906},
pp. \bfpage{694}--\blpage{711}
(\byear{2016}).
\doiurl{10.1007/978-3-319-46475-6_43}
\end{bchapter}
\endbibitem

\bibitem[\protect\citeauthoryear{Wang et~al.}{2003}]{ms-ssim}
\begin{bchapter}
\bauthor{\bsnm{Wang}, \binits{Z.}},
\bauthor{\bsnm{Simoncelli}, \binits{E.}},
\bauthor{\bsnm{Bovik}, \binits{A.}}:
\bctitle{Multiscale structural similarity for image quality assessment}.
In: \bbtitle{Conference Record of the Asilomar Conference on Signals, Systems and Computers},
vol. \bseriesno{2},
pp. \bfpage{1398}--\blpage{1402}
(\byear{2003}).
\doiurl{10.1109/ACSSC.2003.1292216}
\end{bchapter}
\endbibitem

\bibitem[\protect\citeauthoryear{He et~al.}{2016}]{resnet}
\begin{bchapter}
\bauthor{\bsnm{He}, \binits{K.}},
\bauthor{\bsnm{Zhang}, \binits{X.}},
\bauthor{\bsnm{Ren}, \binits{S.}},
\bauthor{\bsnm{Sun}, \binits{J.}}:
\bctitle{Deep residual learning for image recognition}.
In: \bbtitle{Conference on Computer Vision and Pattern Recognition (CVPR)},
pp. \bfpage{770}--\blpage{778}
(\byear{2016}).
\doiurl{10.1109/CVPR.2016.90}
\end{bchapter}
\endbibitem

\bibitem[\protect\citeauthoryear{Ioffe and Szegedy}{2015}]{bn}
\begin{bchapter}
\bauthor{\bsnm{Ioffe}, \binits{S.}},
\bauthor{\bsnm{Szegedy}, \binits{C.}}:
\bctitle{Batch normalization: Accelerating deep network training by reducing internal covariate shift}.
In: \bbtitle{32nd International Conference on Machine Learning (ICML)},
vol. \bseriesno{37},
pp. \bfpage{448}--\blpage{456}
(\byear{2015}).
\doiurl{10.5555/3045118.3045167}
\end{bchapter}
\endbibitem

\bibitem[\protect\citeauthoryear{Nair and Hinton}{2010}]{Nair2010RectifiedLU}
\begin{bchapter}
\bauthor{\bsnm{Nair}, \binits{V.}},
\bauthor{\bsnm{Hinton}, \binits{G.}}:
\bctitle{Rectified linear units improve restricted boltzmann machines}.
In: \bbtitle{International Conference on Machine Learning (ICML)},
vol. \bseriesno{27},
pp. \bfpage{807}--\blpage{814}
(\byear{2010}).
\doiurl{10.5555/3104322.3104425}
\end{bchapter}
\endbibitem

\bibitem[\protect\citeauthoryear{Gordon et~al.}{2014}]{Gorden2014}
\begin{barticle}
\bauthor{\bsnm{Gordon}, \binits{B.}},
\bauthor{\bsnm{Blazey}, \binits{T.}},
\bauthor{\bsnm{Benzinger}, \binits{T.}}:
\batitle{Regional variability in alzheimer’s disease biomarkers}.
\bjtitle{Future Neurology}
\bvolume{9},
\bfpage{131}--\blpage{134}
(\byear{2014})
\doiurl{10.2217/FNL.14.9}
\end{barticle}
\endbibitem

\bibitem[\protect\citeauthoryear{Mueller et~al.}{2005}]{adni}
\begin{barticle}
\bauthor{\bsnm{Mueller}, \binits{S.G.}},
\bauthor{\bsnm{Weiner}, \binits{M.W.}},
\bauthor{\bsnm{Thal}, \binits{L.J.}},
\bauthor{\bsnm{Petersen}, \binits{R.C.}},
\bauthor{\bsnm{Jack}, \binits{C.}},
\bauthor{\bsnm{Jagust}, \binits{W.}},
\bauthor{\bsnm{Trojanowski}, \binits{J.Q.}},
\bauthor{\bsnm{Toga}, \binits{A.W.}},
\bauthor{\bsnm{Beckett}, \binits{L.}}:
\batitle{The alzheimer's disease neuroimaging initiative}.
\bjtitle{Neuroimaging clinics of North America}
\bvolume{15},
\bfpage{869}--\blpage{877}
(\byear{2005})
\doiurl{10.1016/j.nic.2005.09.008}
\end{barticle}
\endbibitem

\bibitem[\protect\citeauthoryear{Jack et~al.}{2010}]{jack2010hypothetical}
\begin{barticle}
\bauthor{\bsnm{Jack}, \binits{C.}},
\bauthor{\bsnm{Knopman}, \binits{D.}},
\bauthor{\bsnm{Jagust}, \binits{W.}},
\bauthor{\bsnm{Shaw}, \binits{L.}},
\bauthor{\bsnm{Aisen}, \binits{P.}},
\bauthor{\bsnm{Weiner}, \binits{M.}},
\bauthor{\bsnm{Petersen}, \binits{R.}},
\bauthor{\bsnm{Trojanowski}, \binits{J.}}:
\batitle{Hypothetical model of dynamic biomarkers of the alzheimer's pathological cascade}.
\bjtitle{Lancet neurology}
\bvolume{9},
\bfpage{119}--\blpage{28}
(\byear{2010})
\doiurl{10.1016/S1474-4422(09)70299-6}
\end{barticle}
\endbibitem

\bibitem[\protect\citeauthoryear{Jack~Jr et~al.}{2008}]{jack2008alzheimer}
\begin{barticle}
\bauthor{\bsnm{Jack~Jr}, \binits{C.R.}},
\bauthor{\bsnm{Bernstein}, \binits{M.A.}},
\bauthor{\bsnm{Fox}, \binits{N.C.}},
\bauthor{\bsnm{Thompson}, \binits{P.}},
\bauthor{\bsnm{Alexander}, \binits{G.}},
\bauthor{\bsnm{Harvey}, \binits{D.}},
\bauthor{\bsnm{Borowski}, \binits{B.}},
\bauthor{\bsnm{Britson}, \binits{P.J.}},
\bauthor{\bsnm{L.~Whitwell}, \binits{J.}},
\bauthor{\bsnm{Ward}, \binits{C.}}, \betal:
\batitle{The alzheimer's disease neuroimaging initiative (adni): Mri methods}.
\bjtitle{Journal of Magnetic Resonance Imaging: An Official Journal of the International Society for Magnetic Resonance in Medicine}
\bvolume{27}(\bissue{4}),
\bfpage{685}--\blpage{691}
(\byear{2008})
\doiurl{10.1002/jmri.21049}
\end{barticle}
\endbibitem

\bibitem[\protect\citeauthoryear{Tzourio-Mazoyer et~al.}{2002}]{parce}
\begin{barticle}
\bauthor{\bsnm{Tzourio-Mazoyer}, \binits{N.}},
\bauthor{\bsnm{Landeau}, \binits{B.}},
\bauthor{\bsnm{DF}, \binits{P.}},
\bauthor{\bsnm{Crivello}, \binits{F.}},
\bauthor{\bsnm{Etard}, \binits{O.N.D.}},
\bauthor{\bsnm{Delcroix}, \binits{N.}},
\bauthor{\bsnm{Mazoyer}, \binits{B.}},
\bauthor{\bsnm{Marc}, \binits{J.}}:
\batitle{Automated anatomical labeling of activations in spm using a macroscopic anatomical parcellation of the mni mri single-subject brain}.
\bjtitle{NeuroImage}
\bvolume{15},
\bfpage{273}--\blpage{89}
(\byear{2002})
\doiurl{10.1006/nimg.2001.0978}
\end{barticle}
\endbibitem

\bibitem[\protect\citeauthoryear{Iglesias et~al.}{2011}]{robex}
\begin{barticle}
\bauthor{\bsnm{Iglesias}, \binits{J.}},
\bauthor{\bsnm{Liu}, \binits{C.-Y.}},
\bauthor{\bsnm{Thompson}, \binits{P.}},
\bauthor{\bsnm{Tu}, \binits{Z.}}:
\batitle{Robust brain extraction across datasets and comparison with publicly available methods}.
\bjtitle{IEEE transactions on medical imaging}
\bvolume{30},
\bfpage{1617}--\blpage{34}
(\byear{2011})
\doiurl{10.1109/TMI.2011.2138152}
\end{barticle}
\endbibitem

\bibitem[\protect\citeauthoryear{Jenkinson et~al.}{2012}]{fsl}
\begin{barticle}
\bauthor{\bsnm{Jenkinson}, \binits{M.}},
\bauthor{\bsnm{Beckmann}, \binits{C.F.}},
\bauthor{\bsnm{Behrens}, \binits{T.E.J.}},
\bauthor{\bsnm{Woolrich}, \binits{M.W.}},
\bauthor{\bsnm{Smith}, \binits{S.M.}}:
\batitle{Fsl}.
\bjtitle{NeuroImage}
\bvolume{62}(\bissue{2}),
\bfpage{782}--\blpage{790}
(\byear{2012})
\doiurl{10.1016/j.neuroimage.2011.09.015}
\end{barticle}
\endbibitem

\bibitem[\protect\citeauthoryear{Paszke et~al.}{2017}]{paszke2017automatic}
\begin{bchapter}
\bauthor{\bsnm{Paszke}, \binits{A.}},
\bauthor{\bsnm{Gross}, \binits{S.}},
\bauthor{\bsnm{Chintala}, \binits{S.}},
\bauthor{\bsnm{Chanan}, \binits{G.}},
\bauthor{\bsnm{Yang}, \binits{E.}},
\bauthor{\bsnm{DeVito}, \binits{Z.}},
\bauthor{\bsnm{Lin}, \binits{Z.}},
\bauthor{\bsnm{Desmaison}, \binits{A.}},
\bauthor{\bsnm{Antiga}, \binits{L.}},
\bauthor{\bsnm{Lerer}, \binits{A.}}:
\bctitle{Automatic differentiation in pytorch}.
In: \bbtitle{NIPS 2017 Workshop on Autodiff}
(\byear{2017})
\end{bchapter}
\endbibitem

\bibitem[\protect\citeauthoryear{Kingma and Ba}{2014}]{adam}
\begin{bchapter}
\bauthor{\bsnm{Kingma}, \binits{D.}},
\bauthor{\bsnm{Ba}, \binits{J.}}:
\bctitle{Adam: A method for stochastic optimization}.
In: \bbtitle{International Conference on Learning Representations (ICLR)}
(\byear{2014}).
\doiurl{10.48550/arXiv.1412.6980}
\end{bchapter}
\endbibitem

\bibitem[\protect\citeauthoryear{Odena et~al.}{2017}]{conditional}
\begin{bchapter}
\bauthor{\bsnm{Odena}, \binits{A.}},
\bauthor{\bsnm{Olah}, \binits{C.}},
\bauthor{\bsnm{Shlens}, \binits{J.}}:
\bctitle{Conditional image synthesis with auxiliary classifier gans}.
In: \bbtitle{34th International Conference on Machine Learning (ICML)},
vol. \bseriesno{70},
pp. \bfpage{2642}--\blpage{2651}
(\byear{2017}).
\doiurl{10.5555/3305890.3305954}
\end{bchapter}
\endbibitem

\bibitem[\protect\citeauthoryear{Wang et~al.}{2020}]{WANG2020e775}
\begin{barticle}
\bauthor{\bsnm{Wang}, \binits{Y.}},
\bauthor{\bsnm{Yin}, \binits{G.}},
\bauthor{\bsnm{Lang}, \binits{J.}},
\bauthor{\bsnm{Wang}, \binits{P.}},
\bauthor{\bsnm{Li}, \binits{J.}}:
\batitle{U-net and gans-based pet synthesis from mri for soft-tissue sarcomas}.
\bjtitle{International Journal of Radiation Oncology*Biology*Physics}
\bvolume{108}(\bissue{3, Supplement}),
\bfpage{775}
(\byear{2020})
\doiurl{10.1016/j.ijrobp.2020.07.226} .
\bcomment{Proceedings of the American Society for Radiation Oncology}
\end{barticle}
\endbibitem

\bibitem[\protect\citeauthoryear{Jung et~al.}{2018}]{pip2pix}
\begin{bchapter}
\bauthor{\bsnm{Jung}, \binits{M.}},
\bauthor{\bsnm{Berg}, \binits{B.}},
\bauthor{\bsnm{Postma}, \binits{E.}},
\bauthor{\bsnm{Huijbers}, \binits{W.}}:
\bctitle{Inferring pet from mri with pix2pix}.
(\byear{2018})
\end{bchapter}
\endbibitem

\bibitem[\protect\citeauthoryear{Sun et~al.}{2019}]{Sun2019DUALGLOWCF}
\begin{botherref}
\oauthor{\bsnm{Sun}, \binits{H.}},
\oauthor{\bsnm{Mehta}, \binits{R.R.}},
\oauthor{\bsnm{Zhou}, \binits{H.H.}},
\oauthor{\bsnm{Huang}, \binits{Z.}},
\oauthor{\bsnm{Johnson}, \binits{S.C.}},
\oauthor{\bsnm{Prabhakaran}, \binits{V.}},
\oauthor{\bsnm{Singh}, \binits{V.}}:
Dual-glow: Conditional flow-based generative model for modality transfer.
2019 IEEE/CVF International Conference on Computer Vision (ICCV),
10610--10619
(2019)
\end{botherref}
\endbibitem

\bibitem[\protect\citeauthoryear{Hu et~al.}{2020}]{gen_gan}
\begin{bbook}
\bauthor{\bsnm{Hu}, \binits{S.}},
\bauthor{\bsnm{Shen}, \binits{Y.}},
\bauthor{\bsnm{Wang}, \binits{S.-Q.}},
\bauthor{\bsnm{Lei}, \binits{B.}}:
\bbtitle{Brain MR to PET Synthesis via Bidirectional Generative Adversarial Network},
pp. \bfpage{698}--\blpage{707}
(\byear{2020}).
\doiurl{10.1007/978-3-030-59713-9_67}
\end{bbook}
\endbibitem

\bibitem[\protect\citeauthoryear{Hu et~al.}{2019}]{gen_adve}
\begin{bchapter}
\bauthor{\bsnm{Hu}, \binits{S.}},
\bauthor{\bsnm{Yuan}, \binits{J.}},
\bauthor{\bsnm{Wang}, \binits{S.}}:
\bctitle{Cross-modality synthesis from mri to pet using adversarial u-net with different normalization}.
In: \bbtitle{2019 International Conference on Medical Imaging Physics and Engineering (ICMIPE)},
pp. \bfpage{1}--\blpage{5}
(\byear{2019}).
\doiurl{10.1109/ICMIPE47306.2019.9098219}
\end{bchapter}
\endbibitem

\bibitem[\protect\citeauthoryear{Zhang et~al.}{2022}]{bpgan}
\begin{barticle}
\bauthor{\bsnm{Zhang}, \binits{J.}},
\bauthor{\bsnm{He}, \binits{X.}},
\bauthor{\bsnm{Qing}, \binits{L.}},
\bauthor{\bsnm{Gao}, \binits{F.}},
\bauthor{\bsnm{Wang}, \binits{B.}}:
\batitle{Bpgan: Brain pet synthesis from mri using generative adversarial network for multi-modal alzheimer’s disease diagnosis}.
\bjtitle{Computer Methods and Programs in Biomedicine}
\bvolume{217},
\bfpage{106676}
(\byear{2022})
\doiurl{10.1016/j.cmpb.2022.106676}
\end{barticle}
\endbibitem

\bibitem[\protect\citeauthoryear{Sawyer et~al.}{2017}]{Sawyer217}
\begin{barticle}
\bauthor{\bsnm{Sawyer}, \binits{R.}},
\bauthor{\bsnm{Rodriguez-Porcel}, \binits{F.}},
\bauthor{\bsnm{Hagen}, \binits{M.}},
\bauthor{\bsnm{Shatz}, \binits{R.}},
\bauthor{\bsnm{Espay}, \binits{A.}}:
\batitle{Diagnosing the frontal variant of alzheimer’s disease: a clinician’s yellow brick road}.
\bjtitle{Journal of Clinical Movement Disorders}
\bvolume{4},
\bfpage{2}
(\byear{2017})
\doiurl{10.1186/s40734-017-0052-4}
\end{barticle}
\endbibitem

\bibitem[\protect\citeauthoryear{Hernández et~al.}{2018}]{Hernandez2018}
\begin{barticle}
\bauthor{\bsnm{Hernández}, \binits{M.}},
\bauthor{\bsnm{Reid}, \binits{S.}},
\bauthor{\bsnm{Mikhael}, \binits{S.}},
\bauthor{\bsnm{Pernet}, \binits{C.}}:
\batitle{Do 2-year changes in superior frontal gyrus and global brain atrophy affect cognition?}
\bjtitle{Alzheimer's \& Dementia: Diagnosis, Assessment \& Disease Monitoring}
\bvolume{10},
\bfpage{706}--\blpage{716}
(\byear{2018})
\doiurl{10.1016/j.dadm.2018.07.010}
\end{barticle}
\endbibitem

\bibitem[\protect\citeauthoryear{Jacobs et~al.}{2017}]{jacobs2018cerebellum}
\begin{barticle}
\bauthor{\bsnm{Jacobs}, \binits{H.}},
\bauthor{\bsnm{Hopkins}, \binits{D.}},
\bauthor{\bsnm{Mayrhofer}, \binits{H.}},
\bauthor{\bsnm{Bruner}, \binits{E.}},
\bauthor{\bsnm{Leeuwen}, \binits{F.}},
\bauthor{\bsnm{Raaijmakers}, \binits{W.}},
\bauthor{\bsnm{Schmahmann}, \binits{J.}}:
\batitle{The cerebellum in alzheimer's disease: Evaluating its role in cognitive decline}.
\bjtitle{Brain : a journal of neurology}
\bvolume{141},
\bfpage{37}--\blpage{47}
(\byear{2017})
\doiurl{10.1093/brain/awx194}
\end{barticle}
\endbibitem

\bibitem[\protect\citeauthoryear{Bau et~al.}{2019}]{GANDissect}
\begin{bchapter}
\bauthor{\bsnm{Bau}, \binits{D.}},
\bauthor{\bsnm{Zhu}, \binits{J.-Y.}},
\bauthor{\bsnm{Strobelt}, \binits{H.}},
\bauthor{\bsnm{Zhou}, \binits{B.}},
\bauthor{\bsnm{Tenenbaum}, \binits{J.B.}},
\bauthor{\bsnm{Freeman}, \binits{W.T.}},
\bauthor{\bsnm{Torralba}, \binits{A.}}:
\bctitle{Gan dissection: Visualizing and understanding generative adversarial networks}.
In: \bbtitle{International Conference on Learning Representations (ICLR)}
(\byear{2019}).
\doiurl{10.48550/arXiv.1811.10597}
\end{bchapter}
\endbibitem

\end{thebibliography}

\end{document}